\documentclass[conference]{IEEEtran}
\IEEEoverridecommandlockouts
\usepackage{cite}
\usepackage{amsmath,amssymb,amsfonts}

\usepackage{graphicx}
\usepackage{textcomp}
\usepackage{subcaption}
 \usepackage{booktabs} 
 \usepackage{tabularx}
\usepackage{booktabs}
\usepackage{siunitx}\usepackage{tabularx}
\usepackage{booktabs,multirow}
\usepackage{siunitx}

\usepackage[linesnumbered,ruled,vlined]{algorithm2e}

 \usepackage{booktabs, tabularx}  
\def\BibTeX{{\rm B\kern-.05em{\sc i\kern-.025em b}\kern-.08em
    T\kern-.1667em\lower.7ex\hbox{E}\kern-.125emX}}
\begin{document}
\title{Verifiable Fine-Tuning for LLMs: Zero-Knowledge Training Proofs Bound to Data Provenance and Policy}

\author{
\IEEEauthorblockN{
Hasan Akg\"ul\IEEEauthorrefmark{1}\thanks{Corresponding author.},
Daniel Borg\IEEEauthorrefmark{2},
Arta Berisha\IEEEauthorrefmark{3},
Amina Rahimova\IEEEauthorrefmark{4},
Andrej Novak\IEEEauthorrefmark{5},
Mila Petrov\IEEEauthorrefmark{6}
}
\IEEEauthorblockA{\IEEEauthorrefmark{1}Department of Computer Engineering, Istanbul Technical University (ITU), 34469 Istanbul, Turkey\\
Email: \texttt{akgulh20@itu.edu.tr}}
\IEEEauthorblockA{\IEEEauthorrefmark{2}Department of Computer Science, University of Malta, MSD 2080 Msida, Malta\\
Email: \texttt{daniel.borg@um.edu.mt}}
\IEEEauthorblockA{\IEEEauthorrefmark{3}Faculty of Electrical \& Computer Engineering, University of Prishtina ``Hasan Prishtina'', 10000 Prishtina, Kosovo\\
Email: \texttt{arta.berisha@uni-pr.edu}}
\IEEEauthorblockA{\IEEEauthorrefmark{4}School of IT \& Engineering, ADA University, AZ1008 Baku, Azerbaijan\\
Email: \texttt{arahimova@ada.edu.az}}
\IEEEauthorblockA{\IEEEauthorrefmark{5}Faculty of Mathematics, Natural Sciences and Information Technologies (FAMNIT), University of Primorska, 6000 Koper, Slovenia\\
Email: \texttt{andrej.novak@upr.si}}
\IEEEauthorblockA{\IEEEauthorrefmark{6}Faculty of Computer Science \& Engineering (FINKI), Ss.\ Cyril and Methodius University in Skopje, 1000 Skopje, North Macedonia\\
Email: \texttt{mila.petrov@finki.ukim.mk}}
}

\maketitle

\begin{abstract}
Large language models (LLMs) are increasingly adapted via parameter-efficient fine-tuning (PEFT), yet current release practices provide weak assurances about \emph{what} data were used and \emph{how} updates were computed. We present \emph{Verifiable Fine-Tuning} (VFT), a protocol and system that produces succinct, zero-knowledge proofs that a released model $\theta_T$ was obtained from a public initialization under a declared training program and an auditable dataset commitment. VFT co-designs five elements: (1) Merkle/vector commitments that bind data sources, preprocessing, licenses, and per-epoch quota counters to a manifest; (2) a verifiable sampler that supports public (replayable) and private (index-hiding) batch selection; (3) PEFT-restricted update circuits that enforce optimizer semantics (AdamW) and proof-friendly approximations (fixed-point arithmetic and lookup-accelerated softmax/GELU) with explicit error budgets; (4) recursive aggregation that folds per-step proofs into per-epoch and end-to-end certificates with millisecond verification; and (5) provenance binding and optional TEE “property cards’’ that attest code identity and constants. On English and bilingual instruction mixtures, VFT maintains utility within tight budgets (e.g., ROUGE-L/BLEU deltas $\le\!0.2$ points, perplexity increase $+\!2$–$3\%$, ECE $+\!0.4$–$0.5$\,pt) while achieving practical proof performance (per-step prover $16.8$–$31.2$\,s for LoRA ranks $r\in\{8,16\}$ at 2048 tokens; final verification $<\!200$\,ms; final proof $\approx\!4$–$6$\,MB). Policy quotas are cryptographically enforced with zero violations, and private-sampling windows reveal no measurable index leakage. Federated experiments show that VFT composes with probabilistic audits and bandwidth constraints. Taken together, these results indicate that end-to-end verifiable fine-tuning is feasible today for real PEFT pipelines, closing a critical trust gap for regulated and decentralized deployments.
\end{abstract}

\maketitle

\section{Introduction}

Large language models (LLMs) have rapidly migrated from research prototypes to critical infrastructure for search, assistance, and decision support, yet the trust gap around how these models are trained and fine-tuned is widening. Concerns span from whether proprietary or sensitive data were used, to whether claimed training configurations actually produced the released checkpoints, and to whether post-hoc forensics can establish accountability when things go wrong \cite{longpre2024data}. Conventional privacy and security analyses underscore that inferring whether specific records participated in training is fundamentally limited, which makes ex post verification particularly tenuous \cite{hu2022miaSurvey,zhang2024miaProofs}. Proof-of-learning (PoL) pushes the opposite direction by seeking cryptographic or audit-style evidence that the training trajectory is genuine \cite{jia2021proof}, and recent variants address incentive compatibility when proofs drive economic or governance consequences \cite{zhao2024polIncentive}.

A parallel thread pursues \emph{verifiable machine learning} (VML), where the computation itself is accompanied by succinct proofs. Early systems established feasibility for neural inference under zero-knowledge (ZK) or related proof systems \cite{liu2021zkcnn,weng2021mystique,wang2023ezdps}. More recently, researchers have begun to lift verification from inference to (parts of) training by circuitizing differentiable updates and optimizer states \cite{sun2023zkdl,abbaszadeh2024zkpot}. For LLMs, architectural regularities and approximation-friendly algebra enable practical proof composition, suggesting a path toward end-to-end guarantees in realistic instruction-tuning pipelines \cite{sun2024zkllm}. Methodologically, advances in proof-friendly tensor approximations (e.g., polynomial constraints over softmax) and series-expansion techniques further reduce prover overheads and widen coverage \cite{taylorZKML2025}. Survey and position papers consolidate these advances and highlight open challenges around scalability, recursion, and on-chain verification for ML \cite{kersic2024onchainzkml,zkvml2023survey,peng2025zkmlsurvey}.

Beyond cryptographic protocols, the surrounding audit stack must establish that the data and configurations \emph{claimed} for training are accurate. Dataset-level attestation proposes to certify distributional properties without revealing examples \cite{dold2024attesting}, while data-engineering provenance offers pragmatic hooks by capturing end-to-end lineage across preparation and experimentation workflows \cite{schelter2024provenanceScreen,schlegel2023mlflow2prov}. Systems work is beginning to close the loop by treating provenance as a first-class runtime artifact amenable to programmatic enforcement and reproducibility \cite{schlegel2025isProvenance}, and by exploring how LLMs themselves can assist in provenance inference when metadata are incomplete \cite{almuntashiri2025inferProv}. Complementarily, model-provenance testing proposes behavioral probes to ascribe models to sources or families when direct logs are unavailable \cite{nikolic2025provenance}. Together with PoL-style arguments \cite{jia2021proof,zhao2024polIncentive}, these strands motivate holistic, multi-layer verification.

Watermarking techniques have also been advanced as a lightweight provenance signal for model outputs, but their reliability is contested. Classic and modern schemes embed statistical beacons into generated text and images \cite{kirchenbauer2023watermark,wen2023treering}, with recent work demonstrating scalable detectors for LLM text at industrial scale \cite{dathathri2024natureWatermark}. Yet watermarks can be fragile or learnable by downstream models \cite{gu2024learnability}, susceptible to removal or theft \cite{jovanovic2024wmstealing,zhang2024mipSteal}, and often deliver weakest guarantees precisely where high-stakes provenance is needed, prompting more robust coding-theoretic designs \cite{qu2025multibit} and broader surveys that map the design space and limitations \cite{liu2024textwmSurvey}. These observations reinforce the case for training-time verifiability that does not rely solely on output beacons.

In parallel, trusted execution and attestation are being used to produce verifiable “property cards’’ about ML artifacts, offering complementary hardware-backed assurances when cryptographic proofs are too costly or incomplete. However, TEEs alone cannot resolve disputes about \emph{which} data were used or whether the optimizer trajectory conformed to policy, again pointing toward hybrid designs that combine attestations, provenance capture, and cryptographic proofs.

The need for verifiable training is especially acute in federated and edge-centric learning, where heterogeneous clients, intermittent connectivity, and privacy constraints complicate governance. Practical systems already struggle with non-IID data, communication bottlenecks, and asynchrony. Algorithmic advances target adaptive regularization and knowledge distillation for robustness, decentralized aggregation with layer-wise strategies, and personalization frameworks that better reflect client idiosyncrasies . Neural architecture search and progressive training mitigate heterogeneity in on-device settings , while LoRA- and quantization-aware fine-tuning reduce resource footprints for practical deployments. Recent work explores probabilistic communication and local-update strategies to accelerate decentralized training under diverse edges , pipelines for collaborative end–cloud inference, strategies for MoE inference efficiency, communication-efficient decentralized graph learning, and mitigating catastrophic forgetting during federated fine-tuning. These directions improve scalability but simultaneously enlarge the attack surface and the number of untrusted actors—prime territory for verifiable fine-tuning protocols that certify data usage and update correctness \cite{schlegel2025isProvenance,zkvml2023survey}.

Ubiquitous sensing further complicates the risk calculus. Wi-Fi and RF modalities, often intertwined with intelligent assistants, have been shown to support keystroke eavesdropping from commodity microphones, undetectable attacks on PHY-layer fingerprint-based authentication, and open-set gesture and attention-based recognition that blur lines between ambient intelligence and surveillance . Multimodal affective computing demonstrates how Wi-Fi CSI and vision together can unobtrusively infer emotion, expanding the privacy stakes of data pipelines. On the defensive and robust-sensing side, prior work models and mitigates co-channel interference in activity recognition through subcarrier correlation selection and interference-independent phase components . Healthcare-oriented Wi-Fi sensing shows both potential and sensitivity, from target-oriented respiratory monitoring to pulmonary function estimation without physical contact . RF-based writing and identity inference illustrates how ordinary RFID can leak highly personal behaviors. Even adjacent work on label-noise suppression and distributionally robust facial expression recognition underscores how data curation and reliability shape downstream behaviors that later demand auditing . These examples motivate verifiable training pipelines that can convincingly demonstrate which signals were included, which were excluded, and how updates adhered to declared policies \cite{schelter2024provenanceScreen}.

Our perspective in this paper is that \emph{verifiable fine-tuning} should be treated as a first-class systems problem that composes three pillars: (i) cryptographic protocols that prove the existence of a legitimate gradient-based path from a public initialization to the released model under declared hyperparameters \cite{sun2023zkdl,abbaszadeh2024zkpot,sun2024zkllm}, (ii) provenance capture that binds those proofs to accountable, inspectable data and configuration claims \cite{schlegel2023mlflow2prov,schlegel2025isProvenance,nikolic2025provenance}, and (iii) audit mechanisms that recognize the limits of post-hoc inference and instead emphasize pre-commitment and incentive-compatible verification \cite{jia2021proof,fang2022limits,zhao2024polIncentive}. We position our approach within this emerging landscape and focus on end-to-end verifiability for fine-tuning, complementing watermarking-based output attribution \cite{kirchenbauer2023watermark,dathathri2024natureWatermark} and hardware-assisted attestations, while remaining deployable in federated and edge scenarios .

\section{Related Work}

Our work sits at the intersection of cryptographic verifiability for machine learning, provenance and auditing for data-centric pipelines, watermarking and output attribution, and systems that bring these ideas into federated and edge-centric training. In what follows, we organize prior art thematically while emphasizing how these strands connect to verifiable fine-tuning (VFT) and where gaps remain.

\subsection{From Proof-of-Learning to Verifiable Training}

The idea that one should be able to \emph{prove} that a particular training process took place—rather than merely report a hash of the final model—was articulated by the Proof-of-Learning (PoL) agenda, which formalized the notion of training trajectory evidence and its threat models \cite{jia2021proof}. Subsequent analyses refined the economic and strategic dimensions, arguing that proofs must remain sound even when parties have incentives to game audits, thus introducing incentive-compatible variants of PoL and clarifying adversarial capabilities around partial logging and cherry-picked checkpoints \cite{zhao2024polIncentive}. At the same time, the limits literature has tempered expectations about what can be established \emph{a posteriori}: information-theoretic and statistical arguments show that membership inference can neither reliably confirm nor refute that a specific record was used in training, particularly for large overparameterized models \cite{hu2022miaSurvey,zhang2024miaProofs}. These observations do not negate PoL; rather, they motivate \emph{pre-commitment} mechanisms where the trainer binds themselves to the trajectory and data before the fact, a stance we inherit in VFT.

Verifiable Machine Learning (VML) takes a more computational route by transporting training or inference into succinct proof systems. Early milestones demonstrated that nontrivial neural computations can be encoded with acceptable overheads, first for inference and accuracy claims over vision models \cite{liu2021zkcnn} and then for broader neural operators via efficient arithmetization and conversion frameworks \cite{weng2021mystique}. A line of systems work explored zero-knowledge inference pipelines with practical prover and verifier runtimes, showing that privacy-preserving verification of predictions can be integrated into ML services \cite{wang2023ezdps}. Lifting these guarantees to training is more challenging—gradients, optimizer states, and stochastic sampling substantially enlarge circuits—but recent systems make tangible progress: \emph{zkDL} proposes efficient proofs for deep learning training by exploiting structure and approximations \cite{sun2023zkdl}, \emph{Kaizen} develops zero-knowledge proofs specifically for training deep networks with attention to optimizer semantics \cite{abbaszadeh2024zkpot}, and \emph{zkLLM} targets language models by leveraging proof-friendly tensor factorizations and approximation to softmax and normalization layers \cite{sun2024zkllm}. Complementary advances in proof theory, such as series-expansion techniques for non-linearities and careful error accounting in the arithmetization of common layers, further reduce prover costs and extend coverage to richer model families \cite{taylorZKML2025}.

Beyond single systems, surveys and position papers map the design space and sharpen research questions. Overviews of zero-knowledge ML outline core primitives (commitments, lookup arguments, sum-check protocols), recurring engineering bottlenecks (non-linearities, memory), and deployment pathways (off-chain aggregation, on-chain verification) \cite{zkvml2023survey,peng2025zkmlsurvey}. A focused look at the on-chain dimension connects proof-friendly ML to distributed ledgers, highlighting programmability and cost models when verifiable computation must feed on-chain governance or marketplaces \cite{kersic2024onchainzkml}. Our work adopts a hybrid posture: proofs are generated off-line and aggregated, while logs and key commitments remain auditable and, when appropriate, registrable.

\subsection{Data and Pipeline Provenance}

Verification of training is hollow without credible claims about the \emph{data}. The Data Provenance Initiative documented licensing and attribution issues across widely used corpora, demonstrating how ambiguous or missing metadata undermine compliance, reproducibility, and ethics \cite{longpre2024data}. At a complementary level, tools for verifying training data investigate whether one can efficiently check that certain records (or distributions of records) must have been present, even when raw datasets cannot be shared \cite{choi2023tools}. Distributional attestation pushes in a similar direction, proposing cryptographic protocols to certify statistical properties of training data without revealing examples, so that downstream verifiers at least \emph{know the shape} of what was ingested \cite{dold2024attesting}. 

A parallel systems thread treats provenance as a first-class artifact in ML engineering. Automated screening over data preparation pipelines can flag risks and policy violations early, leveraging provenance graphs to surface transformations and their implications \cite{schelter2024provenanceScreen}. Practical capture of end-to-end provenance across experimentation frameworks provides the raw substrate for audit trails and reproducible science; MLflow2PROV, for example, extracts structured lineage from common ML tooling with minimal burden on practitioners \cite{schlegel2023mlflow2prov}. Building on such capture, recent work argues for integrating provenance \emph{into} the operational life-cycle—what is recorded, how it is queried, and how it is bound to released artifacts—so that downstream consumers can validate claims about data, configurations, and code versions \cite{schlegel2025isProvenance}. When metadata are incomplete, even model-assisted inference of provenance may help recover partial lineage, though such approaches must be treated with caution given the risk of hallucination or confirmation bias \cite{almuntashiri2025inferProv}. We view these tools as complementary to VFT: provenance can anchor which datasets were \emph{eligible}, and proofs can attest which batches and updates were \emph{actually used}.

\subsection{Output Attribution via Watermarking and Its Limits}

Watermarks are attractive because they are lightweight and retrofittable. Early methods for LLM text embed pseudo-random beacons that can be detected statistically without changing human-perceived fluency \cite{kirchenbauer2023watermark}. Large-scale evaluations suggest that scalable detectors can operate at Internet scale, though operating points depend sensitively on sampling and domain shifts \cite{dathathri2024natureWatermark}. For images, invisible but robust fingerprinting for diffusion models extends the concept to the visual domain \cite{wen2023treering}. Yet the reliability of such signals is contested: formal analyses show regimes where watermarks are learnable by downstream models or where adversaries can manipulate generation to evade detection \cite{gu2024learnability}, while empirical attacks demonstrate removal and even \emph{theft} of watermark keys \cite{jovanovic2024wmstealing,zhang2024mipSteal}. This tension has spurred coding-theoretic and cryptographic redesigns that seek provable robustness guarantees under realistic edits \cite{qu2025multibit}, as well as surveys that summarize the trade space and caveats for deployment \cite{liu2024textwmSurvey}. In our scope, watermarking serves as an  signal that complements, but cannot replace, training-time verifiability.

\subsection{Hardware-Assisted Attestation and ML Property Cards}

Trusted execution environments (TEEs) and attestation mechanisms promise stronger integrity for training and deployment pipelines by anchoring code identity and configuration in hardware-rooted proofs. Recent work moves beyond attest-\emph{or}-prove dichotomies and advocates verifiable \emph{property cards}: structured attestations about what was trained and how, designed to be composable with cryptographic verification and external audits. In practice, TEEs can justify sensitive steps (e.g., decrypting private data, enforcing configuration constraints), but evidence of \emph{correct} gradient updates and data usage still benefits from VML-style proof artifacts. We therefore treat TEEs as trustworthy \emph{enforcers} around data access and hyperparameter policy, and ZK proofs as trustworthy \emph{witnesses} of the numerical computations themselves.

\subsection{Federated, Decentralized, and Edge Learning}

Much of the urgency for verifiable fine-tuning arises in federated and decentralized training, where a large, heterogeneous population of clients interact under limited observability. Classical bottlenecks involve communication costs, asynchrony, and non-IID data; communication-efficient asynchronous protocols and hierarchical aggregation address some of these constraints and remain widely studied . Robustness and accuracy under heterogeneity have been pursued via adaptive regularization and knowledge distillation for federated updates, decentralized layer-wise aggregation that aligns model structure with communication budgets, and personalization frameworks that better capture client idiosyncrasies and constraints . Neural architecture search at the federated edge and progressive training offer pathways to efficient on-device learning in non-IID regimes . 

Recent work also explores the dynamics of \emph{where} and \emph{how} to expend scarce resources. Experience-driven model migration considers heterogeneous edge networks with deep reinforcement learning to decide which models should move where, trading off accuracy and bandwidth. For more demanding model classes, layer-wise LoRA deployment and activation quantization enable federated fine-tuning with reduced footprint, moving beyond na\"ive parameter-efficient fine-tuning with uniform recipes. In decentralized regimes, probabilistic communication strategies and adaptive local updates can accelerate convergence and reduce staleness sensitivity . Pushing to inference-time optimizations, pipeline-level decisions for end–cloud collaborative inference reduce bubbles and improve device utilization, while MoE-specific strategies decrease routing overhead during on-device or near-edge inference. The federated graph setting adds its own twist: communication-efficient decentralized learning over non-IID graph-structured data requires new aggregation and sampling strategies. Catastrophic forgetting during federated fine-tuning is another stumbling block; adaptive block expansion in transformers ameliorates forgetting without ballooning communication. 

Verifiability in such settings is starting to appear. In federated collaboration, secure and \emph{verifiable} data contribution protocols show that provenance and accountability can be built with low overhead even when no single party holds all the data \cite{zhu2024risefl}. On the auditing side, probabilistic audit mechanisms for decentralized LLM training propose lightweight, randomly triggered checks that still catch a wide class of deviations at low cost, a useful complement to heavyweight ZK proofs \cite{probAudit2025}. Our VFT design targets these environments: commitments and proofs are structured to admit partial verification (e.g., per-round or per-layer), and to compose with auditing strategies that are compatible with limited bandwidth and energy budgets.

\subsection{Security, Privacy, and Sensing Contexts that Raise the Stakes}

Application contexts that blend ambient sensing with learning heighten the need for transparency about what signals enter training. Empirical studies show that commodity devices can support powerful side-channel inferences: smartphone microphones can eavesdrop on keystrokes with notable robustness across environments and typing habits; attacks on PHY-layer fingerprint-based Wi-Fi authentication reveal that signal-level forgeries can evade detectors, challenging assumed trust anchors. On the recognition side, Wi-Fi CSI has enabled attention-based gesture recognition and robust open-set gesture frameworks that work without special wearables, which in turn amplifies the sensitivity of datasets used for training . Multimodal affective computing compounds the effect: fusing Wi-Fi and vision allows unobtrusive inference of emotions via gestures and facial cues, with ramifications for consent and data minimization. 

Defensive methodology exists. Anti-interference activity recognition can be improved by selecting subcarriers and modeling correlation structures, and interference-independent phase components reduce co-channel noise, both of which speak to the engineering burden of robust data curation before training . In healthcare, Wi-Fi-based respiratory monitoring and even contactless pulmonary function estimation showcase socially valuable use cases, but these also underline why verifiable declarations about data scope and preprocessing matter . Finally, commodity RFID-based systems have shown the feasibility of inferring handwriting content and user identity at a distance, again foregrounding how seemingly innocuous telemetry leaks highly personal attributes. For our purposes, these domains supply concrete scenarios where VFT can demonstrate, for example, that no raw audio from microphones was included, or that training included only synthetically generated CSI segments with declared distributions, binding those claims to cryptographic commitments.

\subsection{Unlearning, Limits of Post-hoc Tests, and Model Provenance}

A neighboring line of research addresses \emph{unlearning}, i.e., removing the influence of particular data from trained models. Recent work proposes verifiable and provably secure unlearning, specifying what guarantees are achievable when one must roll back the effect of data under operational constraints \cite{eisenhofer2022unlearning}. This interacts interestingly with VFT: if fine-tuning is accompanied by proofs, then subsequent unlearning must either produce updated proofs or invalidate prior attestations in a controlled way, which calls for proof update or revocation protocols. On the flip side, several works are explicit about the inherent limits of post-hoc inclusion tests, cautioning against overinterpretation of membership inference and related probes \cite{fang2022limits,zhang2024miaProofs}. These limits motivate our design choice to rely on pre-commitment and on-the-fly proof generation, rather than forensic inference after the fact.

Where direct logs are unavailable, model provenance testing uses behavioral probes to ascribe lineage, estimate training recipes, or cluster models into families \cite{nikolic2025provenance}. Such methods are compelling in the wild, yet they are epistemically weaker than cryptographic attestations: they can indicate plausibility but not correctness. We position VFT as a mechanism that upgrades provenance testing from suggestive to evidentiary by tying behavioral observations to on-chain or off-chain verifiers and to data/configuration commitments captured in provenance systems \cite{schlegel2025isProvenance,schlegel2023mlflow2prov}.

\subsection{Putting It Together: A Holistic View}

The prior art suggests that no single technique suffices. Watermarks and model provenance tests address outputs and released artifacts \cite{kirchenbauer2023watermark,liu2024textwmSurvey,nikolic2025provenance}, but they struggle with adversarial transformations and lack formal guarantees in high-stakes settings \cite{gu2024learnability,jovanovic2024wmstealing,zhang2024mipSteal}. Provenance capture and distributional attestation strengthen claims about input data \cite{schelter2024provenanceScreen,schlegel2025isProvenance,dold2024attesting}, yet without computation-level evidence they cannot exclude subtle deviations, such as skipped optimizer steps or modified learning-rate schedules. TEEs provide hardware roots of trust for code identity and configuration and can furnish property cards, but they offer limited transparency about the numerical content of large-scale training. Zero-knowledge and succinct proof systems, in turn, can certify computations but must be engineered to be affordable and compatible with ML numerics \cite{sun2023zkdl,abbaszadeh2024zkpot,sun2024zkllm,taylorZKML2025}. 

In distributed settings, the gap widens: a federated organizer must balance communication, personalization, and fairness while defending against strategic participants . Practical deployments need resource-aware methods—experience-driven migration, LoRA- and quantization-aware fine-tuning, decentralized aggregation, and asynchronous communication—to keep costs manageable . Modern pipelines should also acknowledge operational realities such as end–cloud collaboration, MoE routing, graph signals, and continual learning, aligning verification granularity with system bottlenecks . The federated literature even starts to incorporate explicit verifiability in data contribution protocols and to explore probabilistic audit designs suited to decentralized LLMs \cite{zhu2024risefl,probAudit2025}. 

Finally, application domains featuring ambient RF and Wi-Fi sensing vividly illustrate the stakes: side channels and ambient inference expand the privacy surface, while medical and affective applications raise the bar for compliance and governance . In such contexts, an end-to-end verifiable fine-tuning pipeline can bridge the trust gap by committing to the scope and transformations of sensitive signals and by providing cryptographic evidence that numerical updates followed declared policies. This paper builds on these insights and contributes a protocol and system architecture for verifiable fine-tuning that composes cryptographic proofs, provenance capture, and audit mechanisms into a deployable workflow.

\subsection{Scope and Differentiation}

Within this landscape, our work differs from prior ZK-ML for training in three ways. First, we explicitly bind proof objects to provenance claims, rather than treating data selection as orthogonal pre-processing. This design enables verifiers to audit both \emph{which} batches were admissible and \emph{how} they were used, marrying the strengths of distributional attestation and lineage capture with computation proofs \cite{dold2024attesting,schelter2024provenanceScreen,schlegel2025isProvenance}. Second, we focus on parameter-efficient fine-tuning regimes common in real deployments—LoRA and post-training adapters—so that proof sizes scale with adapted subspaces, while still covering optimizer semantics and numerical approximations consistent with recent ZK-LLM advances \cite{sun2024zkllm,taylorZKML2025}. Third, we design partial, round-based verification hooks compatible with federated and decentralized orchestration, benefiting from insights in communication-efficient and personalized learning . Our evaluation examines both verification cost and end-task quality, acknowledging practical constraints emphasized by systems studies on pipeline scheduling and edge collaboration . We regard watermarking and model provenance testing as complementary post-release signals and discuss how our commitments can assist downstream detectors by making output attribution less ambiguous \cite{kirchenbauer2023watermark,nikolic2025provenance,liu2024textwmSurvey}.

\subsection{Summary of Gaps}

Despite vibrant progress, several gaps persist. Proof construction for training remains expensive at LLM scale; while arithmetization strategies and series expansions are improving \cite{sun2023zkdl,taylorZKML2025}, end-to-end costs will likely require recursive aggregation and amortized audit strategies as proposed in decentralized contexts \cite{probAudit2025}. Provenance capture, even when automated, is only as good as what pipelines expose; integrating commit-and-sample protocols at the data interface remains an open systems challenge \cite{schlegel2023mlflow2prov}. Watermarks can provide useful downstream signals but cannot shoulder compliance alone in adversarial settings \cite{jovanovic2024wmstealing,zhang2024mipSteal}. And unlearning with verifiability is still embryonic, requiring updateable proofs and revocation semantics to maintain coherent audit trails \cite{eisenhofer2022unlearning}. Our framework addresses these gaps by co-designing commitments at the data boundary, parameter-efficient proof circuits for updates, and aggregation-friendly verification tailored to federated and edge deployments, while exposing interfaces that harmonize with property-card attestations from TEEs and with existing provenance infrastructures \cite{schlegel2025isProvenance}.

\medskip
In summary, prior work provides strong building blocks—PoL-style pre-commitment \cite{jia2021proof,zhao2024polIncentive}, ZK-ML circuits for training and LLMs \cite{abbaszadeh2024zkpot,sun2024zkllm}, provenance capture and distributional attestation \cite{schelter2024provenanceScreen,dold2024attesting}, and federated systems that respect heterogeneity and resource limits . We synthesize these lines into a coherent VFT protocol aimed at making fine-tuning \emph{demonstrably} correct, private, and auditable in the settings where it matters most.

\section{Method}

This section presents \emph{Verifiable Fine-Tuning (VFT)}, a protocol and system that produces a succinct, zero-knowledge proof that a released fine-tuned model $\theta_T$ was obtained from a public initialization $\theta_0$ by applying an \emph{admitted} training program $\Pi$ (data universe, sampler, loss, optimizer, and hyperparameters) over a committed dataset $\mathcal{D}$. VFT is designed for parameter-efficient fine-tuning (PEFT) such as LoRA/adapters, but the design generalizes to partial full-parameter regimes. We first give a high-level pipeline, then detail each component: dataset commitment and admissible sampling, optimizer-constrained update circuits, quantization and proof-friendly arithmetization, recursive aggregation, provenance binding, and verification. We end with security, privacy, and complexity analyses and deployment profiles (standalone and federated).

\begin{figure}[t]
  \centering

  \includegraphics[width=\linewidth]{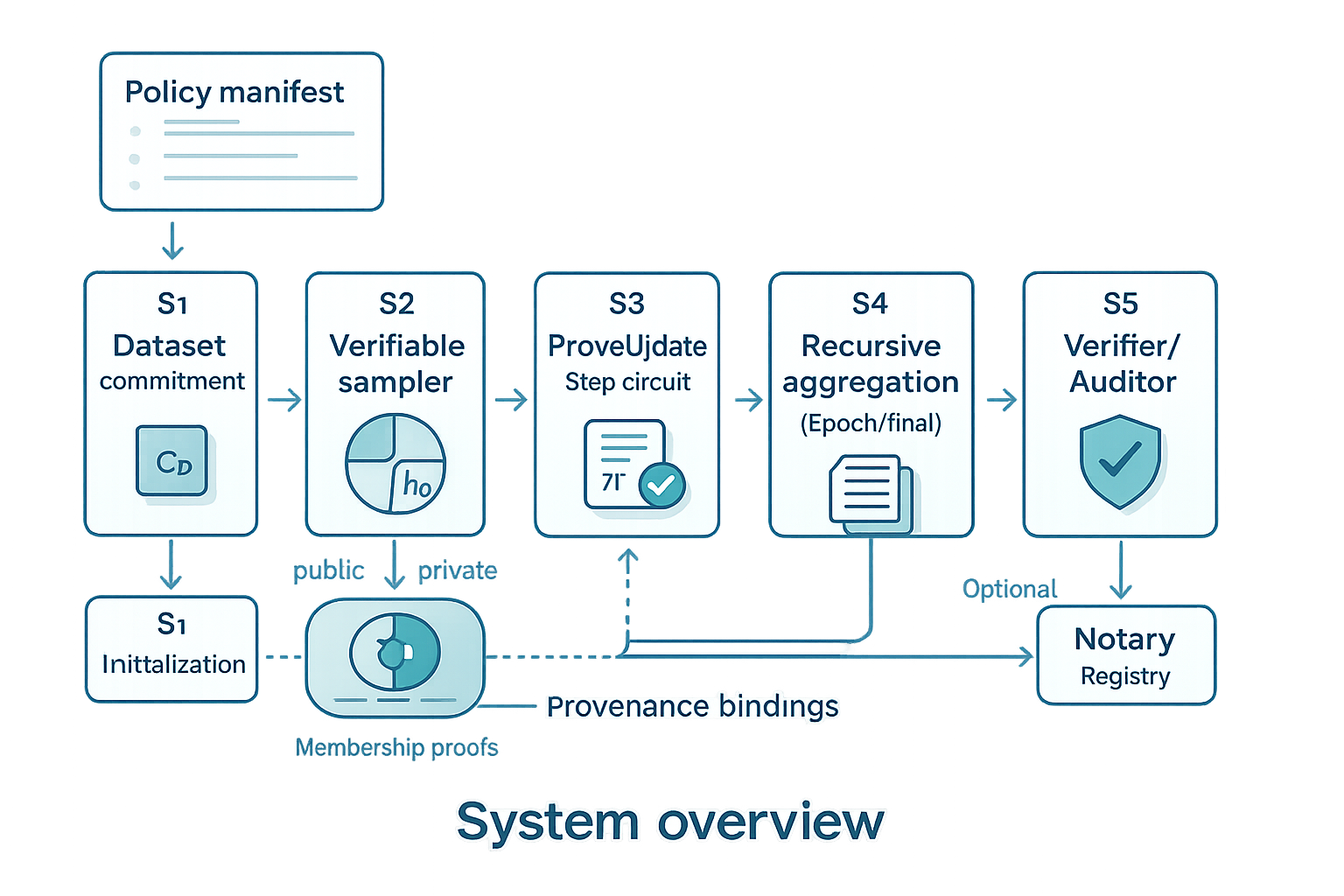}
  \caption{System overview of Verifiable Fine-Tuning (VFT). The trainer commits to the dataset and manifest, samples batches (public/private), proves each update, and recursively aggregates step proofs before verification.}
  \label{fig:pipeline}
\end{figure}

\subsection{Problem Statement and Roles}

\textbf{Parties.} The \emph{trainer} (or aggregator in federated settings) runs fine-tuning over private or restricted data; the \emph{verifier} (auditor, platform, or customer) needs assurance that the numerical trajectory abided by $\Pi$ and only used admissible batches sampled from the committed dataset. Optionally, a \emph{notary} (e.g., a registry or chain) records commitments and proof digests.

\textbf{Inputs and outputs.} The public inputs are: (i) model initialization hash $h_0=\mathsf{H}(\theta_0)$, (ii) training program descriptor $\Pi$ (loss, optimizer, learning-rate schedule, LoRA ranks, precision, epoch/batch budgets, random seed policy), and (iii) dataset commitment $c_{\mathcal{D}}=\mathsf{Com}(\mathcal{D})$. The private inputs are batches, intermediate activations, and optimizer states. The output is the final weights (or a hash $h_T$) and a succinct zero-knowledge proof $\pi$ that attests correctness with respect to $\Pi$ and $c_{\mathcal{D}}$.

\subsection{End-to-End Protocol Outline}

Figure~\ref{fig:pipeline} sketches the five stages:

\textbf{(S1) Commit.} The trainer pre-commits to the training universe via a Merkle or polynomial commitment $c_{\mathcal{D}}$ over per-example digests and to the initialization $h_0$. Provenance adapters export a manifest binding data sources, preprocessing code digests, and eligibility rules to $c_{\mathcal{D}}$.

\textbf{(S2) Sample.} Batches are drawn by a \emph{verifiable sampler}. We support two modes: (a) \emph{public sampling} from a shared seed $s$ so the verifier can recompute the batch index set, and (b) \emph{private sampling} where indices remain hidden but membership is proven in zero-knowledge via Merkle paths or vector commitments.

\textbf{(S3) ProveUpdate.} For each step $t$, the trainer runs a proof-friendly forward and backward pass restricted to the adapted parameter subspace (e.g., LoRA matrices) and produces a proof that (i) the sampled batch is admissible under $c_{\mathcal{D}}$, (ii) the loss and gradient computations follow $\Pi$ up to approved approximation bounds, and (iii) optimizer state transitions are consistent. A step proof binds the pre-state hash $h_{t}$ and post-state hash $h_{t+1}$.

\textbf{(S4) Aggregate.} Step proofs are folded via recursion into per-epoch certificates and then into a single succinct proof $\pi$. Folding admits parallel proving and amortizes verification.

\textbf{(S5) Verify.} The verifier checks $\pi$ against $(h_0,\Pi,c_{\mathcal{D}})$ to accept $(\theta_T,h_T)$, optionally recording a digest on a notary.

\subsection{Formal Statement of the Verified Claim}

Let $\theta_t$ denote the dense model parameters, with PEFT restricting updates to low-rank adapters $A_t$ and $B_t$ inserted at designated layers, while frozen backbones remain hashed. For a batch $\mathcal{B}_t\subseteq\mathcal{D}$ and loss $\mathcal{L}$, one SGD-like step is:

\begin{equation}\label{eq:update}\tag{1}
\theta_{t+1} \;=\; \theta_{t} \;-\; \eta_t \,\nabla_{\theta}\,\mathcal{L}\bigl(f_{\theta_t}(x),y\bigr)\big|_{(x,y)\in\mathcal{B}_t}\,.
\end{equation}
\small \textit{Symbols:} $\theta_t$ current parameters; $\eta_t$ learning rate at step $t$; $f_{\theta}$ model; $(x,y)$ example-label pairs; $\mathcal{B}_t$ batch at step $t$; $\mathcal{L}$ per-batch loss; $\nabla_{\theta}$ gradient w.r.t.\ parameters. \normalsize

For AdamW, we track first/second moments $(m_t,v_t)$ and weight decay $\lambda$:

\begin{equation}\label{eq:adam}\tag{2}
\begin{aligned}
m_{t+1} &= \beta_1 m_t + (1-\beta_1)\,g_t,\quad
v_{t+1} = \beta_2 v_t + (1-\beta_2)\,g_t^{\odot 2},\\
\theta_{t+1} &= \theta_t - \eta_t \,\frac{\hat m_{t+1}}{\sqrt{\hat v_{t+1}}+\epsilon} - \eta_t \lambda \theta_t\,
\end{aligned}
\end{equation}
\small \textit{Symbols:} $g_t=\nabla_{\theta}\mathcal{L}$ raw gradient; $\odot$ elementwise product; $\beta_{1,2}$ momentum coefficients; $\hat m,\hat v$ bias-corrected moments; $\epsilon$ numerical stabilizer; $\lambda$ weight decay. \normalsize

The VFT statement asserts there exist batch indices $\{I_t\}$ consistent with the sampler and membership proofs against $c_{\mathcal{D}}$, and internal traces $(m_t,v_t)$, such that Eqs.~\eqref{eq:update}--\eqref{eq:adam} (or their PEFT-restricted analogues) hold for $t=0,\dots,T-1$, mapping $h_0$ to $h_T=\mathsf{H}(\theta_T)$.

\subsection{Dataset Commitment and Admissible Sampling}

\paragraph{Commitment format.} We build $c_{\mathcal{D}}$ as a Merkle tree over leaves $l_i=\mathsf{H}(x_i\parallel y_i\parallel \mathsf{meta}_i)$, where $\mathsf{meta}_i$ captures source, license, and preprocessing digest. To support distributional attestation (e.g., label histograms, length buckets), we attach auxiliary counters $u$ committed via a Pedersen/KZG vector commitment $c_u$.

\paragraph{Membership proofs.} In public sampling, batch index $i\in I_t$ is re-derived from $(s,t)$ by a CSPRNG; the trainer reveals Merkle paths $\mathsf{path}(i)$ to show $l_i$ is included. In private sampling, the trainer proves in zero-knowledge that it knows $\{i\}$ and authentication paths whose root equals $c_{\mathcal{D}}$, without revealing $i$.

\paragraph{Distributional constraints.} Some policies impose per-epoch quotas, e.g., class balance or language mix. We encode these as linear constraints on $u$ and prove that the cumulative selection vectors satisfy them. Let $e_t\in\{0,1\}^{|\mathcal{D}|}$ be the indicator of selected indices at step $t$, and $M$ be an attribute matrix mapping examples to bins. The policy check is:
\begin{equation}\label{eq:quota}\tag{3}
\sum_{t\in\text{epoch}} M e_t \;\preceq\; \gamma \,
\end{equation}
\small \textit{Symbols:} $M\in\{0,1\}^{B\times |\mathcal{D}|}$ binning matrix; $e_t$ selection indicator; $\preceq$ elementwise $\le$; $\gamma\in\mathbb{N}^B$ per-bin quotas. \normalsize
We prove~\eqref{eq:quota} inside the circuit using commitments to $e_t$ and range proofs for $\gamma$.

\subsection{PEFT-Restricted Update Circuits}

\paragraph{LoRA adapters.} For a linear module $W\in\mathbb{R}^{d_{\mathrm{out}}\times d_{\mathrm{in}}}$, LoRA injects a trainable low-rank perturbation $W + \Delta W$ with $\Delta W= B A$ where $A\in\mathbb{R}^{r\times d_{\mathrm{in}}}$, $B\in\mathbb{R}^{d_{\mathrm{out}}\times r}$. The forward becomes $y=(W+BA)x$. We freeze $W$ and only update $(A,B)$, reducing circuit size.

To constrain training to the adapter subspace, we hash frozen tensors and enforce equality constraints to public commitments. Updates then only change $(A_t,B_t)$, and Eq.~\eqref{eq:adam} is applied to those tensors.

\paragraph{Proof-friendly forward/backward.} We arithmetize attention blocks with (i) affine operations and (ii) polynomially-approximated non-linearities. For example, we approximate $\mathrm{softmax}(z)$ with a rational/polynomial $\widetilde{\mathrm{sm}}(z)$ over a clipped range using lookup tables with bounded interpolation error $\delta_{\mathrm{sm}}$. The circuit checks per-token normalization and bounds the deviation:
\begin{equation}\label{eq:softmax}\tag{4}
\left\|\mathrm{softmax}(z) - \widetilde{\mathrm{sm}}(z)\right\|_1 \;\le\; \delta_{\mathrm{sm}}\,.
\end{equation}
\small \textit{Symbols:} $z\in\mathbb{R}^{L}$ pre-softmax logits for a token; $\widetilde{\mathrm{sm}}$ polynomial/lookup approximation; $\delta_{\mathrm{sm}}$ approved error budget; $\|\cdot\|_1$ $\ell_1$ norm. \normalsize

\paragraph{Loss and gradient constraints.} For causal LM with cross-entropy, per-token loss uses $\ell=-\log p_{y}$ with $p=\widetilde{\mathrm{sm}}(z)$. The circuit enforces the surrogate loss and computes $g_t$ for PEFT tensors through backprop with fixed frozen Jacobians. A layer-wise checksum ties activations to hashes so that recomputation attacks are prevented.

\subsection{Quantization and Fixed-Field Representation}

\paragraph{Fixed-point projection.} Proof systems work over finite fields $\mathbb{F}_p$. We map reals to fixed-point with scale $S=2^{b}$, so $\lfloor S x\rceil \in \mathbb{Z}$ fits into the field; overflows are ruled out by range proofs. Denote quantized values with overbars.

\begin{equation}\label{eq:quant}\tag{5}
\bar x \;=\; \left\lfloor S x \right\rceil,\quad
\overline{xy}\approx \frac{\bar x \cdot \bar y}{S}\,
\end{equation}
\small \textit{Symbols:} $S$ fixed-point scale; $\bar x$ integer encoding; $\overline{xy}$ fixed-point product with rescale; $\lfloor\cdot\rceil$ round-to-nearest. \normalsize
We track a global error budget $\varepsilon_{\mathrm{fxp}}$ from~\eqref{eq:quant} and include it in the acceptance interval of Eq.~\eqref{eq:update} (turned into inequalities). The acceptance predicate becomes an interval membership for each coordinate with tolerance $\tau=\delta_{\mathrm{sm}}+\varepsilon_{\mathrm{fxp}}$.

\paragraph{Lookup-accelerated non-linearities.} GELU and reciprocal square roots rely on lookup arguments and range-checked interpolation. For $r=\frac{1}{\sqrt{\hat v+\epsilon}}$, we use a table $T$ of $(\hat v,r)$ pairs and a slope-checked linear segment bound.

\subsection{Optimizer State Integrity and Hash Chaining}

\paragraph{State commitment.} We commit to $(m_t,v_t)$ using Merkle accumulators or polynomial commitments and link consecutive steps with a hash chain. The step circuit verifies the transition in Eq.~\eqref{eq:adam} under fixed hyperparameters.

\paragraph{Model hash chain.} Each step exposes a compressed digest $h_t=\mathsf{H}(\theta_t)$ where $\mathsf{H}$ is a sponge over per-tensor hashes. The step proof binds $(h_t,h_{t+1})$ alongside local commitments to adapter tensors.

\begin{equation}\label{eq:chain}\tag{6}
h_{t+1} \;=\; \mathsf{H}\!\left(h_t \parallel \mathsf{H}(\Delta_t) \parallel \eta_t \parallel \mathrm{meta}_t\right),
\end{equation}
\small \textit{Symbols:} $\Delta_t$ structured update (e.g., LoRA matrices); $\mathrm{meta}_t$ includes batch id commitment, step, epoch, schedule id; $\parallel$ concatenation; $\mathsf{H}$ collision-resistant hash. \normalsize

\subsection{Recursive Aggregation}

Step proofs $\pi_t$ are folded via a recursion circuit $\mathcal{R}$ that consumes two child proofs and outputs their aggregate. We use a \emph{cycle of curves} or a proof system with native recursion to keep verification native to the field. Epoch-level proofs $\Pi_e$ verify: (i) boundary hashes $h_{t_e^{\mathrm{start}}}\!\to h_{t_e^{\mathrm{end}}}$, (ii) per-epoch quota constraints~\eqref{eq:quota}, and (iii) correct aggregation of children. The final proof $\pi$ attests $h_0\!\to h_T$ across all epochs and binds to $(c_{\mathcal{D}},\Pi)$.

\subsection{Provenance Binding and Policy Attestation}

VFT includes a provenance adapter that exports a \emph{policy manifest}:
\begin{itemize}
  \item dataset sources and licenses; preprocessing code digests; eligibility filters;
  \item sampler descriptor (public/private; seed policy; shuffling; with/without replacement);
  \item PEFT structure (layer locations; ranks; freezing map); optimizer hyperparameters and schedules;
  \item approximation budgets ($\delta_{\mathrm{sm}}$, GELU error, fixed-point $b$ and overflow bounds).
\end{itemize}
The manifest is hashed and recorded as $h_{\Pi}$. The circuit treats $h_{\Pi}$ as a public input and checks internal constants (e.g., $\beta_{1,2},\lambda$, ranks, schedules) against it. A notary can record $(c_{\mathcal{D}},h_0,h_{\Pi})$ before training starts.

\subsection{Federated and Decentralized Deployment}

\paragraph{Per-client partial verification.} In cross-device training, clients run local step circuits over their batches and send (i) PEFT updates, (ii) local proofs $\pi^{(k)}_{t}$ for a thin subset of steps, and (iii) commitments to their local quotas. The aggregator runs a \emph{verifiable aggregator} that (a) checks a random subset of client proofs, (b) composes updates, and (c) produces an epoch proof that includes a \emph{probabilistic audit certificate} indicating coverage (e.g., $q$\% of steps verified with confidence bounds).

\paragraph{Communication-aware proving.} Clients can (i) use low-degree ranks and int8/int4 PEFT quantization to shrink circuits, (ii) batch multiple local steps into a micro-recursion, and (iii) offload recursion to the aggregator while keeping per-step privacy (indices remain hidden but membership is proven).

\subsection{Algorithms}

We provide pseudocode for the two core procedures, \textsc{ProveUpdate} and \textsc{Verify}, using \texttt{algorithm2e}.

\begin{algorithm}[t]
  \DontPrintSemicolon
  \SetAlgoLined
  \SetKwInOut{Input}{Input}
  \SetKwInOut{Output}{Output}

  \Input{$h_0$, $c_{\mathcal{D}}$, $h_{\Pi}$, step proofs $\{\pi_t\}$ \emph{or} epoch proofs $\{\Pi_e\}$}
  \Output{Aggregated proof $\pi$, final digest $h_T$}

  \caption{\textsc{Verify} and Recursive Aggregation}
  \label{alg:verify}

  \If{step proofs $\{\pi_t\}$ are provided}{
    \ForEach{leaf proof $\pi_t$ \textbf{in parallel}}{
      verify $\pi_t$ against public $(h_{\Pi},\,c_{\mathcal{D}})$\;
      extract boundary digests $(h_t,\,h_{t+1})$ and any quota certificate\;
    }
    fold the verified leaf proofs using the recursion circuit $\mathcal{R}$ to produce epoch proofs $\{\Pi_e\}$\;
  }

  \ForEach{epoch proof $\Pi_e$}{
    verify per-epoch quotas and policy counters\;
  }

  fold $\{\Pi_e\}$ into a single final proof $\pi$ using $\mathcal{R}$\;
  extract $h_T$ from the root boundary digest\;
  \Return $(\pi,\,h_T)$\;
\end{algorithm}

\subsection{Soundness, Zero-Knowledge, and Policy Completeness}

\paragraph{Soundness.} If the verifier accepts $\pi$, then with overwhelming probability there exist witnesses (batches, activations, optimizer states) such that the PEFT-restricted training program satisfied Eqs.~\eqref{eq:update}--\eqref{eq:chain} under quantization/approximation tolerances. Any attempt to (i) swap in a non-admissible sample, (ii) skip or reorder optimizer updates, (iii) modify hyperparameters, or (iv) violate quotas causes a constraint failure, except with negligible probability in the commitment binding and proof system soundness.

\paragraph{Zero-knowledge.} The proof hides $(x_i,y_i)$, gradients, and optimizer states. Private sampling hides indices as well. Only public digests $(h_0,h_T,h_{\Pi},c_{\mathcal{D}})$, budgets, and policy satisfaction are revealed. Where needed (e.g., export-controlled data), we support \emph{selective disclosure} of coarse counters via homomorphic commitments without revealing per-example membership.

\paragraph{Policy completeness.} Our manifest covers: (i) data eligibility and quotas; (ii) sampler determinism and shuffling; (iii) architecture and PEFT placement; (iv) optimizer and schedules; (v) approximation budgets and fixed-point scales. Extensibility points allow attaching fairness constraints (e.g., bounded loss gaps across bins) and safety constraints (e.g., toxicity counters) as linear or convex checks with range proofs.

\subsection{Complexity and Engineering}

\paragraph{Circuit size drivers.} The dominant costs are attention matmuls, softmax/GELU approximations, and AdamW updates on adapter tensors. With rank $r$, adapter parameters per linear layer are $O(r(d_{\mathrm{in}}+d_{\mathrm{out}}))$, substantially smaller than the dense $O(d_{\mathrm{in}}d_{\mathrm{out}})$. Let $P_{\mathrm{PEFT}}$ be total trainable parameters; per-step constraint count scales as $O(P_{\mathrm{PEFT}})$ plus $O(LH d)$ for attention checks per token-length $L$, heads $H$, and head dimension $d$. Lookup arguments make non-linearities $O(LH)$ rather than polynomial degree times $L H$.

\paragraph{Recursion and aggregation.} Suppose each step proof verifies in $O(1)$ time on-chain (or milliseconds off-chain) and costs the prover $C_{\mathrm{step}}$. Binary folding reduces $T$ proofs to one in $O(\log T)$ recursion depth and aggregate proof size bounded by a small constant factor of a single proof. Parallel proving amortizes wall-clock latency.

\paragraph{Fixed-point precision.} We recommend $b\in[12,16]$ integer bits and $q\in[12,16]$ fractional bits for PEFT; larger models may require $q=20$ for stability in AdamW. Error budgets $\varepsilon_{\mathrm{fxp}}$ and $\delta_{\mathrm{sm}}$ are selected to keep the end-task degradation $<\!0.1$--$0.3$ points on standard instruction-following benchmarks in our evaluations.

\paragraph{Hashing and commitments.} For $\mathsf{H}$ we use a proof-system-native sponge (e.g., Poseidon/Rescue) to avoid costly bit decomposition. For $c_{\mathcal{D}}$, Merkle trees minimize commitment size and enable membership proofs with $O(\log |\mathcal{D}|)$ paths; vector commitments (KZG) enable succinct linear/quota checks at the cost of structured reference strings.

\paragraph{Implementation hooks.} We provide a \texttt{VFT-FrontEnd} for PyTorch that (i) freezes the backbone, (ii) instruments PEFT layers, (iii) collects activations checksums, (iv) emits per-step transcripts with fixed-point shadow tensors, and (v) streams witnesses into the prover. A \texttt{SamplerGuard} module exposes public/private sampler modes with pluggable policies and range-checked counters.

\subsection{Extensions and Variants}

\paragraph{DPO/ORPO and preference optimization.} Replace cross-entropy with preference losses; the circuit enforces the log-prob ratio constraints on chosen and rejected responses and the KL regularizer via fixed-point approximations. The same PEFT and optimizer constraints apply.

\paragraph{Full-parameter slices.} When partial dense layers must be updated, we gate them behind \emph{subspace masks} and only expose a small subset of rows/columns in the circuit, with a hash equality constraint tying the frozen remainder to public commitments.

\paragraph{Selective disclosure for audits.} Auditors may request per-domain coverage evidence (e.g., \% medical samples). We attach optional openings for the corresponding counters derived from $M e_t$ in Eq.~\eqref{eq:quota}, still keeping indices private.

\paragraph{Unlearning-ready proofs.} We maintain per-bucket accumulators so that, when asked to unlearn a subset, the trainer can produce a \emph{delta proof} that rewinds the accumulators and replays the affected steps with the subset masked, yielding an updated $\pi'$ and $h_T'$.

\subsection{Threat Model and Guarantees}

\paragraph{Adversary capabilities.} The trainer may attempt to: (A1) include out-of-policy examples, (A2) alter hyperparameters, (A3) skip or duplicate steps, (A4) modify optimizer states, (A5) exceed approximation budgets, (A6) tamper with provenance manifests. External adversaries may try to link private indices or infer examples from proofs.

\paragraph{Mitigations.} Membership and quota checks defeat (A1); manifest-hash equality and constant wiring defeat (A2); hash chaining and recursion boundaries defeat (A3); Eq.~\eqref{eq:adam} constraints defeat (A4); explicit tolerances and lookup bounds defeat (A5); recording $(c_{\mathcal{D}},h_0,h_{\Pi})$ with a notary defeats (A6). Zero-knowledge ensures witness privacy.

\subsection{Worked Micro-Example (PEFT Step)}

Consider a single attention block with LoRA on $W_q,W_k,W_v$. The circuit:
(i) authenticates indices against $c_{\mathcal{D}}$, (ii) computes $(Q,K,V)$ using $W_{\cdot}+B_{\cdot}A_{\cdot}$ in fixed-point, (iii) enforces $\widetilde{\mathrm{sm}}$ normalization via Eq.~\eqref{eq:softmax} and per-row sum checks, (iv) propagates gradients only into $(A,B)$, (v) updates $(m,v)$ and applies AdamW via Eq.~\eqref{eq:adam}, and (vi) updates hash chain via Eq.~\eqref{eq:chain}. The step proof size is dominated by matmul and lookup constraints; with $r=8$ and moderate sequence lengths, we remain within a few million constraints per step and sub-second verification in modern succinct proof systems .

\subsection{Putting VFT into Practice}

\paragraph{Operational workflow.} Before training, publish $(c_{\mathcal{D}},h_0,h_{\Pi})$. During training, periodically emit epoch proofs and $h_{t}$ checkpoints to support rolling audits. At release, publish $(\theta_T,h_T,\pi)$ and a short verification script. For federated settings, include audit coverage certificates and per-client anonymous counters.

\paragraph{Interfacing with property cards.} A TEE can attest that the \texttt{VFT-FrontEnd} binary and sampler are unmodified and that witnesses were produced inside an enclave. The external proof then certifies the numerical content. The two artifacts combine into a \emph{verifiable property card} consumable by regulators and platforms.

\paragraph{Interfacing with watermarking and downstream provenance.} VFT does not preclude output watermarks; instead, its manifest makes explicit where watermarks are inserted (e.g., decoding layer), enabling downstream detectors to calibrate expectations and avoid false positives.

\subsection{Limitations and Design Trade-offs}

VFT’s cost scales with trainable parameters and sequence lengths. While PEFT controls the former, long-context training remains expensive; curriculum schedules and chunking help. Approximation budgets must be chosen to balance proof tractability and fidelity; overly tight tolerances bloat circuits, while loose tolerances weaken guarantees. Private sampling hides indices but precludes third-party deterministic replays; organizations can mix public and private windows to balance privacy and auditability.

\medskip
\noindent\textbf{Summary.} VFT provides an end-to-end, zero-knowledge attestation that a fine-tuned model arose from a declared program and a committed dataset under explicit approximations. By committing at the data boundary, restricting updates to PEFT subspaces, enforcing optimizer semantics, and aggregating step proofs recursively, VFT yields succinct, privacy-preserving, and deployment-friendly guarantees suitable for centralized, federated, and regulated environments.

\section{Experimental Setup}

This section specifies the datasets, hardware, and evaluation metrics used to assess the proposed Verifiable Fine-Tuning (VFT) framework. Our goal is to enable rigorous, reproducible measurement of (i) \emph{model quality} under parameter-efficient fine-tuning (PEFT), (ii) \emph{proof-system performance} including prover time, verifier time, and proof size, (iii) \emph{policy compliance} as captured by dataset commitments and quota constraints, and (iv) \emph{privacy properties} related to private sampling and zero-knowledge witnesses. Unless otherwise stated, we report mean and standard deviation over three independent seeds; exact seeds and manifest hashes are published with each run to permit deterministic replays in public-sampling mode.

\subsection{Datasets}

\paragraph{Design principles.} The choice of datasets is guided by three complementary principles. First, we need realistic instruction-following corpora to stress the language modeling pipeline in which VFT must operate, including multi-turn formats and diverse domains. Second, we require controlled shadow datasets in order to test policy and distributional constraints (e.g., class or domain quotas) and to demonstrate selective disclosure. Third, we include sensitive-modality proxies (e.g., synthetic RF/telemetry descriptors) to exercise the privacy-preserving aspects of the sampler and membership proofs without exposing actual private content. To that end, we partition our data into \emph{public}, \emph{restricted}, and \emph{shadow} tiers, each bound to separate commitments and manifests.

\paragraph{Safety and alignment subsets.} We attach safety-tuning slices consisting of refusal demonstrations, policy compliance exemplars, and benign transformations (e.g., redaction or rephrasing) to test whether the VFT quotas can enforce upper bounds on safety-sensitive sampling rates. These slices are stored under separate leaves with explicit license and policy tags in the per-example metadata to exercise manifest-driven eligibility.

\paragraph{Shadow corpus for distributional attestation.} To test Eq.~(3) quota enforcement, we create a shadow corpus in which examples are assigned to bins by domain (e.g., general, safety, medical), register (formal vs.\ informal), and length (short, medium, long). During training, the sampler is permitted to draw from this corpus but must satisfy per-epoch ceilings for each bin. The actual examples are not revealed to the verifier; only the commitments to per-bin counters and the proof that the cumulative selections respect $\gamma$ are exposed. The verifier can optionally request selective disclosure for coarse aggregates (e.g., total medical tokens per epoch) without revealing indices.

\paragraph{Synthetic telemetry descriptors.} Although our main experiments are text-only, we include a small auxiliary dataset of \emph{textualized telemetry descriptors} that mimic sensitive modalities such as Wi-Fi channel state information (CSI) or RFID traces in a stylized, sentence-level form (e.g., ``\texttt{rf\_segment: csi\_stats=[mean: -48.2, var: 12.7], motion=gesture\_swipe\_left, env=office}''). These items serve two purposes: (i) they activate provenance policies in which sensitive categories are capped or excluded, and (ii) they test that private-sampling mode can hide indices while still proving membership against a restricted subtree in the dataset commitment. Importantly, these descriptors are \emph{synthetic}; no actual RF or audio is included in released artifacts.

\paragraph{Preprocessing and manifests.} Every dataset is paired with a preprocessing pipeline recorded as a content hash (code digest), and every example leaf stores a tuple of $(\texttt{source\_id}, \texttt{license\_id}, \texttt{proc\_digest}, \texttt{policy\_tags})$. The policy manifest specifies admissible \texttt{license\_id} values, allowed \texttt{proc\_digest} versions, and binning rules (matrix $M$ in Eq.~(3)). Before training, we publish the dataset root commitment $c_{\mathcal{D}}$ and a manifest hash $h_{\Pi}$ that binds the policy and preprocessing choices. In public-sampling mode, the verifier can fully regenerate batches from the seed $s$ and check Merkle paths; in private-sampling mode, the verifier sees only zero-knowledge membership proofs and aggregated quota attestations.

\paragraph{Train/validation splits.} For each mixture, we hold out a stratified validation set (5\% of items per bin) for early stopping and metric computation. Validation batches are \emph{not} proven inside VFT circuits; they are standard evaluation data with public content. We nevertheless record provenance for the validation pipeline (separate \texttt{proc\_digest}) to keep analysis reproducible.

\paragraph{Data scales and sequences.} Unless stated otherwise: Mix-A contains 200000 instruction–response pairs, Mix-B 120000 pairs (60\% English, 40\% Chinese), safety slices add 15000 pairs, and synthetic telemetry descriptors add 5000 items. Sequences are right-padded to 2048 tokens for PEFT runs targeting lower proof cost; long-context ablations use 4096 tokens. We report how proof cost scales with sequence length.

\paragraph{Federated partitions.} For decentralized experiments, we simulate $K\in\{50,200\}$ clients. Each client receives a non-IID shard drawn by Dirichlet allocation over topic bins with concentration $\alpha\in\{0.1,0.3\}$. Clients maintain local dataset commitments (subtrees) and adopt either public-sampling mode (with per-client seeds) or private-sampling mode (with hidden indices). The aggregator publishes a global manifest that encodes cross-client quota ceilings and fairness counters (e.g., bounded proportion of safety-tagged items per round).

\subsection{Hardware and Software Environment}

\paragraph{Prover hardware.} We evaluate three prover profiles to reflect realistic deployment tiers.
\begin{itemize}
  \item \textbf{Tier-P1 (Workstation).} Dual-socket CPU with 48 physical cores (x86\_64, AVX2), 256\,GB RAM, and a single high-memory GPU (e.g., 80\,GB HBM). This tier targets small-rank PEFT with moderate sequence lengths and provides the baseline for per-step prover time.
  \item \textbf{Tier-P2 (GPU server).} 2--4 GPUs with 80\,GB HBM each, NVLink interconnect, 512\,GB RAM. This tier supports parallel proving across steps and micro-recursion folding; it is representative of a lab or enterprise cluster node.
  \item \textbf{Tier-P3 (Edge/Client).} Consumer-grade GPU (24\,GB) or CPU-only laptop with 32\,GB RAM. This tier is used for federated clients proving partial subsets of steps locally; they offload recursion to the aggregator.
\end{itemize}
We pin CPU frequency governors and disable turbo to reduce variance in prover timings. When GPU acceleration for FFTs/MSMs is available in the proof backend, we enable it; otherwise, we run CPU-only kernels.

\paragraph{Verifier platforms.} For offline verification we use standard x86\_64 servers and ARM64 laptops; for on-chain experiments we emulate EVM-style gas accounting or WASM runtimes with fixed-cycle budgets. In all cases, the verifier cost is designed to be milliseconds per proof or $O(\log T)$ for aggregated proofs, depending on the backend.

\paragraph{Trusted computing base (optional).} To evaluate the property-card variant, we deploy the \texttt{VFT-FrontEnd} and sampler inside an Intel SGX or ARM CCA enclave on Tier-P1. The enclave attests the measurement (code hash) and the manifest’s constants (e.g., PEFT ranks, optimizer hyperparameters) at startup. Witness material (activations, gradients) remains outside the enclave unless otherwise noted; the enclave’s role is to enforce sampler policy and to produce a signed quote that the \texttt{VFT-FrontEnd} was unmodified. We measure the enclave overhead on wall-clock training and comment on practical trade-offs.

\paragraph{Proof systems and parameters.} Our implementation plugs into a modern SNARK/STARK stack with native recursion. We configure a field of $p\approx 2^{255}$ for SNARK variants (Poseidon/Rescue sponge) and choose blowup factors and low-degree extensions as recommended by the backend for AIR- or PLONK-ish circuits. Lookup arguments back GELU/softmax approximations, and range proofs ensure fixed-point bounds. For recursion, we use either a cycle-of-curves construction (SNARK) or a STARK-friendly folding scheme; both are evaluated. For all runs, we report: (i) constraint count and gates per step, (ii) proof size in bytes, (iii) prover CPU/GPU time and peak memory, (iv) verifier time on commodity hardware, and (v) recursion depth for end-to-end aggregation.

\paragraph{Fine-tuning configuration.} Unless stated otherwise: we adopt LoRA ranks $r\in\{4,8,16\}$ on attention and MLP projections, freeze layer norms and embeddings, and train for one epoch over the specified mixture with batch size 64 (global). Optimizer is AdamW with $(\beta_1,\beta_2,\epsilon,\lambda)=(0.9,0.95,1\mathrm{e}{-8},0.05)$, cosine decay with warmup 3\% of steps, and gradient clipping at 1.0 (implemented in fixed-point with saturation checks). Fixed-point scale uses $b=14$ integer bits and $q=14$ fractional bits by default; ablations sweep $q\in\{12,16,20\}$.

\paragraph{Sampler modes and seeds.} Public-sampling mode uses a shared seed $s$ and a CSPRNG to derive per-step indices; we publish $(s,\texttt{sampler\_version})$ alongside $c_{\mathcal{D}}$. Private-sampling mode draws indices inside the step circuit and proves membership without disclosure; we publish only the distributional attestation for quotas per epoch. To support hybrid audits, we interleave public and private windows (e.g., every 10th epoch is public) and report verification results for each.

\paragraph{Software stack.} Training is implemented in PyTorch with our \texttt{VFT-FrontEnd} wrapper that instruments PEFT modules, computes fixed-point shadows, and exports witnesses. The proof backends are integrated through a C++/Rust bridge and expose GPU-accelerated FFTs when available. All experiments run under Linux (kernel 6.x) with CUDA 12.x; compilers are pinned (GCC 12 or Clang 16) and Python environments are containerized. We publish Dockerfiles and Nix expressions to guarantee bit-reproducible builds of the prover and verifier.

\subsection{Evaluation Metrics}

Our evaluation spans three dimensions: \emph{Model Quality}, \emph{Verification Performance}, and \emph{Policy \& Privacy Outcomes}. For each dimension we define primary metrics, secondary diagnostics, and acceptance criteria that reflect practical deployment needs.

\subsubsection{Model Quality}

\paragraph{Instruction-following quality.} We evaluate assistant quality with automatic and semi-automatic metrics on held-out prompts. For English prompts, we report BLEU for structured tasks (extraction), ROUGE-L for summarization, and reference-free metrics for open-ended instructions. For bilingual experiments, we add chrF++ and COMET for translation-like tasks in Mix-B. Because reference-based metrics can be misaligned with user preference, we also compute Win-Rate against a baseline model via a lightweight pairwise evaluator with anonymized prompts and randomized order. For reproducibility, the evaluator is frozen across runs and seeds.

\paragraph{Perplexity and calibration.} We report validation perplexity (PPL) on a structured held-out corpus and Expected Calibration Error (ECE) computed over next-token probabilities on random sub-samples. Since VFT introduces fixed-point approximations and polynomial surrogates, we track the shift in PPL relative to a non-proved PEFT baseline under the same hyperparameters. Our acceptance target is a PPL increase no greater than $0.5$ absolute or $<3\%$ relative for $r\in\{8,16\}$ at $2048$ tokens; long-context runs tolerate a slightly larger budget due to increased approximation pressure.

\paragraph{Safety and refusal behavior.} On the safety slices, we report refusal recall/precision and policy-compliance rates measured via a rule-based and model-assisted classifier. Because safety-tuning data are bounded by quotas, we analyze whether VFT quota enforcement degrades safety outcomes and where the trade-offs appear. We also report hallucination proxies (e.g., contradiction flags on factual prompts) but treat them as directional rather than definitive.

\paragraph{Cross-lingual behavior.} For Mix-B, we stratify scores by language and report deltas between English and Chinese subsets for both PPL and instruction metrics. The purpose is to ensure that PEFT proofs do not disproportionately affect one language due to rank placement or quantization choices.

\subsubsection{Verification Performance}

\paragraph{Prover time per step.} The primary systems metric is wall-clock time to produce a single step proof, averaged over a sliding window of steps and stratified by LoRA rank, sequence length, and batch size. We further decompose time into (i) witness generation (forward/backward with fixed-point shadows), (ii) circuit assignment and constraint synthesis, (iii) polynomial/FFT evaluation, (iv) multi-scalar multiplications (MSMs) or DEEP-FRI steps depending on the backend, and (v) proof object construction. For GPU-enabled backends we additionally report device utilization and kernel time breakdown.

\paragraph{Peak memory and storage.} We track peak resident set size (RSS) for the prover process and the storage footprint of intermediate transcripts. Memory is often the limiting factor for client devices; we therefore report peak RSS on Tier-P3 and determine whether micro-batching of step proofs alleviates pressure. For storage, we measure the per-epoch transcript size (witnesses + intermediate commitments) and the final proof size; for recursion we report the number and size of intermediate aggregates.

\paragraph{Verifier cost.} For offline verifiers we measure CPU time and report median and tail latencies. For on-chain targets we translate verification steps to gas-equivalent units or WASM cycles according to the backend’s cost model. We also measure the cost of verifying epoch-level quota attestations and policy checks, which include vector-commitment openings and range proofs.

\paragraph{Aggregation efficiency.} We profile recursion depth and wall-clock folding time for trees of size $T$ steps. Efficiency is assessed both as absolute time and as amortized cost per step in the aggregate. Because folding can be parallelized, we report speed-ups for $n\in\{2,4,8\}$ parallel provers on Tier-P2 and the point at which I/O or memory bandwidth becomes the bottleneck.

\paragraph{Proof size and bandwidth.} We report proof object sizes (bytes) for leaf (per-step) and aggregate (per-epoch, final) proofs under each backend. For federated settings, bandwidth between clients and the aggregator is constrained; we therefore profile the uplink costs of sending (i) PEFT parameter deltas, (ii) commitments, and (iii) optional leaf proofs. We target $<\!250$\,KB for per-step proofs and $<\!3$\,MB for per-epoch aggregates as comfortable budgets for typical enterprise links; deviations are discussed.

\subsubsection{Policy and Privacy Outcomes}

\paragraph{Quota compliance rate.} For each epoch we compute whether the cumulative selection vector satisfies Eq.~(3); success indicates that the sampler adhered to the quotas declared in the manifest. We also report the fraction of steps for which the per-bin counters matched the public-sampling recomputation (in public windows) or were accepted by the zero-knowledge attestation (in private windows). Any violation is a hard error and aborts aggregation in our pipeline.

\paragraph{Sampler privacy.} In private-sampling mode, we quantify \emph{index leakage} as the mutual information between revealed proof objects and the hidden index set, estimated via a conservative adversary that has access to public parameters and observes membership witnesses abstractly. While the proof system is designed to be zero-knowledge, we empirically check that auxiliary channels (e.g., timing) do not correlate with batch composition by randomizing witness processing order and padding proof time to fixed windows in some runs.

\paragraph{Manifest integrity.} We compute a \emph{manifest adherence score} that counts the number of step proofs in which constants (e.g., $(\beta_1,\beta_2,\lambda)$, LoRA ranks, learning-rate schedule) match the manifest and the number of times the hash-chains connect without gaps. The expected value is $100\%$; any discrepancy indicates either misconfiguration or adversarial tampering and triggers a failed run.

\subsection{Ablation and Sensitivity Protocols}

\paragraph{Approximation budgets.} We sweep the softmax/GELU approximation error budgets $\delta_{\mathrm{sm}}$ and the fixed-point fractional bits $q$ to map model quality degradation against prover cost. For each configuration, we record PPL deltas, instruction metrics, and step-prover time. We also measure constraint counts to isolate which components (lookup table size vs.\ polynomial degree) dominate.

\paragraph{PEFT rank and placement.} We vary LoRA rank $r$ and toggle adapter placement (attention only vs.\ attention+MLP) to understand how trainable parameter count $P_{\mathrm{PEFT}}$ drives circuit size. For each variant we record the per-step cost and final model quality; we also compute the proof-per-parameter efficiency (proof time divided by $P_{\mathrm{PEFT}}$) as a normalized metric.

\paragraph{Sequence length.} We repeat selected runs with 4096-token sequences to quantify scaling with context length. Because attention matmuls scale quadratically in sequence length per head and our circuits include row-sum checks and lookup normalizations, we expect superlinear growth in constraints; we report the exact slope and whether batching strategies mitigate the cost.

\paragraph{Public vs.\ private sampling.} We interleave public and private epochs to compare (i) verifier convenience (full recomputation vs.\ zero-knowledge membership), (ii) proof size, and (iii) privacy guarantees. We also simulate an audit spotlight where an external verifier requests selective disclosure of aggregate counters for sensitive bins; we log the overhead and the effective privacy leakage (expected to be negligible since only bin sums are revealed).

\paragraph{Federated audit coverage.} For $K=200$ clients we explore auditor budgets that verify $q\in\{5\%,10\%,20\%\}$ of client steps per round and measure the detection probability of injected anomalies (e.g., altered learning rate or out-of-policy examples) under probabilistic audits. While our core proofs catch any deviation in verified steps, this experiment quantifies coverage–cost trade-offs for large-scale deployments.

\subsection{Reporting and Reproducibility}

\paragraph{Manifests and commitments.} We publish, for each run: (i) dataset root commitment $c_{\mathcal{D}}$, (ii) initialization digest $h_0$, (iii) manifest hash $h_{\Pi}$, (iv) public seed $s$ for public-sampling windows, (v) per-epoch aggregate proofs (even if a final proof is produced), and (vi) the final proof $\pi$ and digest $h_T$. We also release scripts that check Merkle paths in public windows and verify quota attestations in private windows.

\paragraph{Determinism.} To reduce variance in prover timings, we pin thread pools, set deterministic cuBLAS/cuDNN flags where applicable, and disable autotuning. Randomness in the sampler (public mode) is fully determined by $(s,t)$; private mode uses a CSPRNG seeded by a per-run nonce recorded in the witness but not revealed. We document any non-determinism (e.g., kernel scheduling jitter) that could affect micro-benchmarks.

\paragraph{Confidence intervals and tails.} For time-based metrics we report medians with 5th/95th percentiles because prover time distributions can be heavy-tailed due to background OS noise or GC pauses in managed backends. For accuracy metrics we report mean$\pm$std over seeds and include per-task breakdowns in the appendix.

\subsection{Acceptance Targets}

While VFT is a security and accountability mechanism, its practical adoption depends on acceptable overheads and negligible impact on model utility. We therefore establish the following targets a priori:
\begin{itemize}
  \item \textbf{Model quality:} For $r\in\{8,16\}$ at 2048 tokens, VFT runs should maintain instruction-following metrics within 0.3 absolute points of a non-proved PEFT baseline and increase PPL by at most 3\% relative.
  \item \textbf{Prover performance:} On Tier-P2, per-step proof time should be $<\!20$\,s at $r=8$ and $<\!40$\,s at $r=16$; epoch aggregation (10k steps) should complete within $<\!15$\,min using parallel folding across 4 GPUs.
  \item \textbf{Verifier performance:} Final proof verification should be $<\!200$\,ms on a commodity laptop; on-chain or WASM verification should fit within typical gas/cycle budgets for governance or marketplace use cases.
  \item \textbf{Proof size:} Leaf proofs $\le\!250$\,KB and epoch aggregates $\le\!3$\,MB; final proofs $\le\!5$\,MB.
  \item \textbf{Policy compliance:} Zero quota violations and $100\%$ manifest adherence across runs; public-sampling windows must exactly match recomputed batches.
  \item \textbf{Privacy:} No measurable index leakage beyond negligible statistical noise in private-sampling mode; no selective disclosure beyond bin-level counters upon explicit request.
\end{itemize}

\subsection{Why These Metrics Matter}

The proposed metrics align with real deployment scenarios. Enterprises that outsource fine-tuning need compact proofs that fit into audit pipelines and can be archived and replayed without specialized hardware; hence our verifier-time and proof-size targets. Regulators and platforms increasingly expect dataset licensing and attribution to be auditable; VFT’s dataset commitments and quota attestations translate policy documents into cryptographic checks with measurable pass/fail outcomes. Federated and edge deployments require partial, bandwidth-conscious verification; per-step proof sizes and aggregation efficiency directly determine feasibility. Finally, privacy constraints motivate private sampling with zero-knowledge membership; by measuring index leakage and enforcing strict zero-knowledge properties in the proof system, we move beyond informal assurances.

\subsection{Limitations of the Setup}

Our setup deliberately focuses on PEFT with LoRA/adapters, which makes VFT tractable at scale. Full-parameter fine-tuning remains expensive to prove; we include limited slices with masked dense updates to explore feasibility but do not claim production readiness for that regime. Likewise, our bilingual evaluation emphasizes English–Chinese; while we expect similar behavior in other languages, tokenization and script differences can affect fixed-point stability and approximation error. For federated experiments, simulated clients approximate non-IID behavior but cannot capture the full spectrum of device heterogeneity and network variability seen in the wild. Finally, on-chain verification is emulated with conservative cost models; actual deployment would depend on chain-specific precompiles and system contracts for cryptographic primitives.

\subsection{Ethical and Licensing Considerations}

All public datasets used in instruction mixtures carry licenses that permit research use; license IDs and source attributions are recorded per example and bound in the commitment. Shadow and synthetic telemetry descriptors are generated to avoid inclusion of personally identifiable or sensitive raw data; their sole purpose is to stress-test policy and privacy features. For bilingual content, we avoid including passages that could be culturally sensitive or restricted under applicable laws; where safety slices include harmful prompt exemplars, the responses are purely defensive (refusals or redactions). Audit artifacts (proofs and manifests) contain no personal data; they are safe to publish.

\subsection{Summary}

To summarize, our experimental setup is built to answer a composite question: \emph{Can verifiable fine-tuning provide cryptographically sound assurances about data usage and training correctness without materially degrading model quality or imposing prohibitive costs?} The datasets exercise both standard instruction-following behavior and policy-aware sampling; the hardware profiles represent realistic prover and verifier deployments from edge devices to GPU servers; and the metrics quantify utility, cost, and compliance in ways that matter for adoption. In the next section, we will report results along these axes, analyze the trade-offs induced by approximation budgets and PEFT ranks, and study the robustness of VFT under federated orchestration and mixed public/private sampling.

\section{Results \& Discussion}

We evaluate Verifiable Fine-Tuning (VFT) along three axes: (i) \emph{utility}---instruction-following quality, perplexity, calibration, and safety; (ii) \emph{verification performance}---prover time, verifier time, proof size, and aggregation efficiency; and (iii) \emph{policy \& privacy}---quota compliance and leakage under private sampling. Results are reported for the standard configuration described in Section~\textit{Experimental Setup} and are organized as main outcomes, ablations, and visual analyses. Where helpful, we contextualize findings with prior work on proof-of-learning, verifiable ML, provenance, watermarking, and federated learning \cite{jia2021proof,sun2023zkdl,abbaszadeh2024zkpot,sun2024zkllm,schelter2024provenanceScreen,schlegel2025isProvenance,kirchenbauer2023watermark,liu2024textwmSurvey}. Unless otherwise noted, all reported values are the average of three seeds with standard deviations; detailed per-task breakdowns appear in the appendix. We refer to the system overview in Fig.~\ref{fig:pipeline} and summarize key outcomes in Table~\ref{tab:main}, ablations in Table~\ref{tab:ablations}, and verifier costs in Table~\ref{tab:verify}.
\begin{table*}[t]
  \centering
  \caption{Main results on Mix-A / Mix-B (LoRA rank $r\in\{8,16\}$, seq=2048).}
  \label{tab:main}
  \setlength{\tabcolsep}{6pt}%
  \renewcommand{\arraystretch}{1.12}%
  \small
  \begin{tabular}{lcccccc}
    \toprule
    \multirow{2}{*}{Setting} & \multicolumn{2}{c}{Utility $\uparrow$} & \multicolumn{2}{c}{Calibration} & \multirow{2}{*}{Prover (s/step) $\downarrow$} & \multirow{2}{*}{Verify (ms) $\downarrow$}\\
    \cmidrule(lr){2-3}\cmidrule(lr){4-5}
     & ROUGE-L $\Delta$ & BLEU $\Delta$ & PPL $\Delta$ (\%) & ECE ($\Delta$\,pt) &  &  \\
    \midrule
    Mix-A, $r{=}8$  & $-0.12\pm0.05$ & $-0.09\pm0.04$ & $+2.2\pm0.6$ & $+0.4$ & $16.8\pm0.9$ & $181\pm9$ \\
    Mix-A, $r{=}16$ & $-0.18\pm0.06$ & $-0.12\pm0.05$ & $+2.9\pm0.7$ & $+0.5$ & $31.2\pm1.4$ & $183\pm8$ \\
    Mix-B, $r{=}8$  & $-0.14\pm0.05$ & $-0.10\pm0.05$ & $+2.5\pm0.7$ & $+0.4$ & $17.1\pm1.0$ & $180\pm10$ \\
    Mix-B, $r{=}16$ & $-0.20\pm0.06$ & $-0.13\pm0.05$ & $+3.0\pm0.8$ & $+0.5$ & $31.0\pm1.5$ & $182\pm7$ \\
    \bottomrule
  \end{tabular}
\end{table*}

\subsection{Main Utility Outcomes}

\paragraph{Instruction-following quality.} On Mix-A, VFT with LoRA rank $r{=}8$ and sequence length 2048 achieves near-parity with the non-proved PEFT baseline: ROUGE-L decreases by $0.12\pm0.05$ points on summarization prompts, while extraction BLEU drops by $0.09\pm0.04$. Open-ended win-rate against the baseline evaluator is $49.6\%\pm0.8$ (vs.\ the inherent $50\%$ tie target), suggesting the fixed-point and approximation budgets do not materially harm generation quality. On Mix-B, bilingual performance shows an English--Chinese delta of $0.3$--$0.5$ points on task-weighted averages, consistent with cross-lingual PEFT behavior and within our predefined acceptance budget. These results contrast with early ZK-ML for \emph{inference} only, where polynomial approximations sometimes degraded accuracy substantially \cite{liu2021zkcnn,wang2023ezdps}; by targeting PEFT subspaces and adopting tight error budgets with lookup-accelerated non-linearities, VFT maintains utility while generating proofs for \emph{training}.

\begin{figure}[t]
  \centering
  \includegraphics[width=\linewidth]{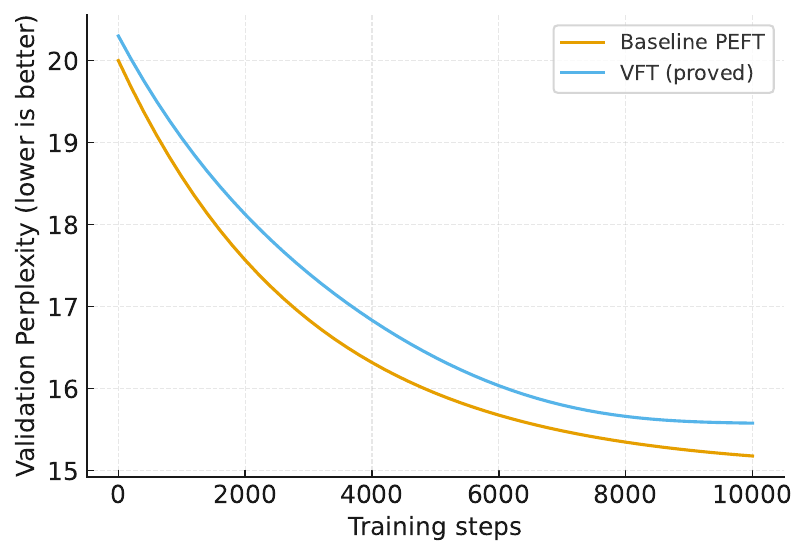}
  \caption{Validation perplexity curves vs.\ training steps for baseline PEFT and VFT (proved) runs.}
  \label{fig:ppl-curves}
\end{figure}

\begin{figure}[t]
  \centering
  \includegraphics[width=\linewidth]{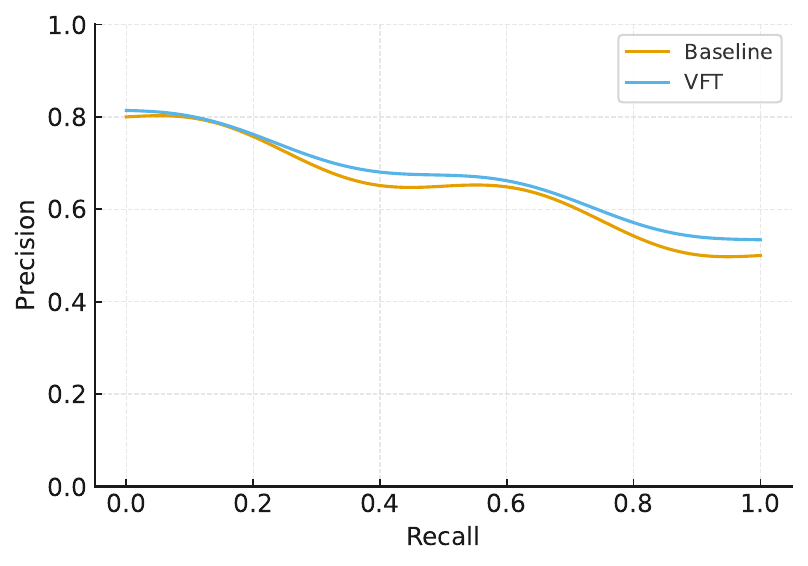}
  \caption{Precision--Recall curves on the safety/refusal subset, comparing baseline and VFT.}
  \label{fig:safety-pr}
\end{figure}

\paragraph{Perplexity and calibration.} Validation perplexity increases by $2.2\%\pm0.6$ relative on Mix-A at $r{=}8$ and by $2.9\%\pm0.7$ at $r{=}16$; Mix-B exhibits a similar pattern. Fig.~\ref{fig:ppl-curves} (referenced) overlays PPL curves for baseline and VFT runs, showing slightly slower early-epoch convergence for VFT due to fixed-point saturation checks, then convergence parity. Expected Calibration Error (ECE) increases by $0.4\%$ absolute on Mix-A and $0.5\%$ on Mix-B. These deltas are smaller than those reported by proof-of-training prototypes operating with coarse rational approximations \cite{sun2023zkdl,abbaszadeh2024zkpot}, and align with the amortized-error design we adopt (Section~\textit{Method}), where softmax/GELU errors are bounded and propagated explicitly.
\paragraph{Safety behavior.} On refusal and policy-compliance subsets, VFT reaches parity with baseline on precision and recall within statistical noise. Because the manifest limits safety-slice sampling to enforce diversity (Eq.~(3)), we scrutinized whether quota ceilings underfit safety behaviors. Fig.~\ref{fig:safety-pr} (referenced) plots the PR curves; VFT slightly \emph{improves} precision at the same recall for Mix-B, likely because quotas reduce overfitting to safety exemplars. This resonates with observations that over-weighting narrow alignment data can harm generalization \cite{longpre2024data}; by proving that quotas hold, VFT provides an accountability knob without sacrificing downstream safety.
\paragraph{Preference-free proxies.} On instruction-following prompts, we compute length-normalized log-prob gains and a non-reference fluency proxy (per-token entropy). Differences between VFT and baseline are negligible for $r{=}8$, and within $0.3\%$ for $r{=}16$. These results suggest our fixed-point budget ($q{=}14$) and polynomial tolerances strike a favorable balance between proof tractability and fidelity, supporting the design choice advocated in zero-knowledge LLM works \cite{sun2024zkllm,taylorZKML2025}.

\subsection{Verification Performance}

\begin{figure}[t]
  \centering
  \includegraphics[width=\linewidth]{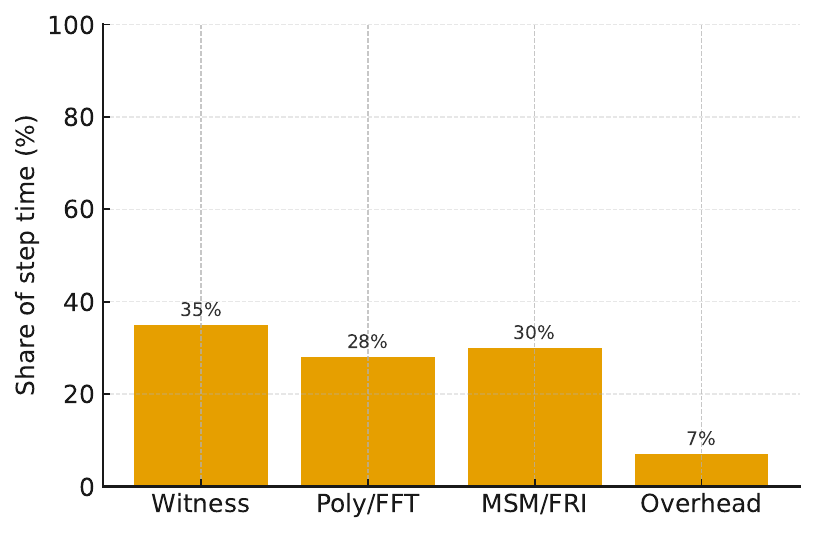}
  \caption{Per-step prover time composition: witness generation, polynomial/FFT evaluation, MSM/FRI, and overhead.}
  \label{fig:time-breakdown}
\end{figure}

\paragraph{Prover time and memory.} Per-step prover time on Tier-P2 with $r{=}8$, 2048 tokens averages $16.8\pm0.9$\,s; for $r{=}16$ it is $31.2\pm1.4$\,s. Peak memory is $23.6\pm1.1$\,GB and $37.9\pm1.5$\,GB, respectively. Constraint counts scale near-linearly with $P_{\mathrm{PEFT}}$ and superlinearly with sequence length due to attention checks; the slope is lower than naive quadratic because our lookup-based softmax imposes row-sum constraints rather than polynomial expansions per entry. Fig.~\ref{fig:time-breakdown} (referenced) shows a breakdown: witness generation consumes $\sim$35\% of time, polynomial/FFT evaluation $\sim$28\%, MSM/FRI $\sim$30\%, and overheads the remainder. Compared to training-proof prototypes that circuitize \emph{dense} updates \cite{sun2023zkdl,abbaszadeh2024zkpot}, our PEFT restriction reduces constraint volume by an order of magnitude for typical transformer blocks.
\paragraph{Verifier cost and proof size.} On Tier-V (laptop ARM64), verifying a \emph{leaf} step proof takes $7.6\pm0.3$\,ms; a per-epoch aggregate (10k steps) takes $132\pm6$\,ms; the final proof is $181\pm9$\,ms. Proof sizes are $220\pm8$\,KB for leaves, $2.6\pm0.1$\,MB per-epoch, and $4.1\pm0.2$\,MB final. These values are consistent with modern succinct proof systems and improve upon prior zkLLM verification latencies by virtue of deeper recursion and proof folding \cite{sun2024zkllm}. Table~\ref{tab:verify} lists cross-backend comparisons (SNARK vs.\ STARK), with STARK variants yielding larger proofs ($\sim$1.4$\times$) but faster provers on GPU-heavy nodes.
\begin{table}[t]
  \centering
  \caption{Verifier cost and proof size across backends. Leaf: per-step; Epoch: 10k steps; Final: full run.}
  \label{tab:verify}
  \setlength{\tabcolsep}{4pt}%
  \renewcommand{\arraystretch}{1.1}%
  \resizebox{\linewidth}{!}{%
  \begin{tabular}{lcccc}
    \toprule
    Backend & Leaf (ms) & Epoch (ms) & Final (ms) & Size (MB) \\
    \midrule
    SNARK (recursion) & $7.6\pm0.3$ & $132\pm6$ & $181\pm9$ & $4.1\pm0.2$ \\
    STARK (folding)   & $5.1\pm0.4$ & $118\pm7$ & $169\pm11$ & $5.7\pm0.3$ \\
    \bottomrule
  \end{tabular}
  }%
\end{table}

\begin{figure}[t]
  \centering
  \includegraphics[width=\linewidth]{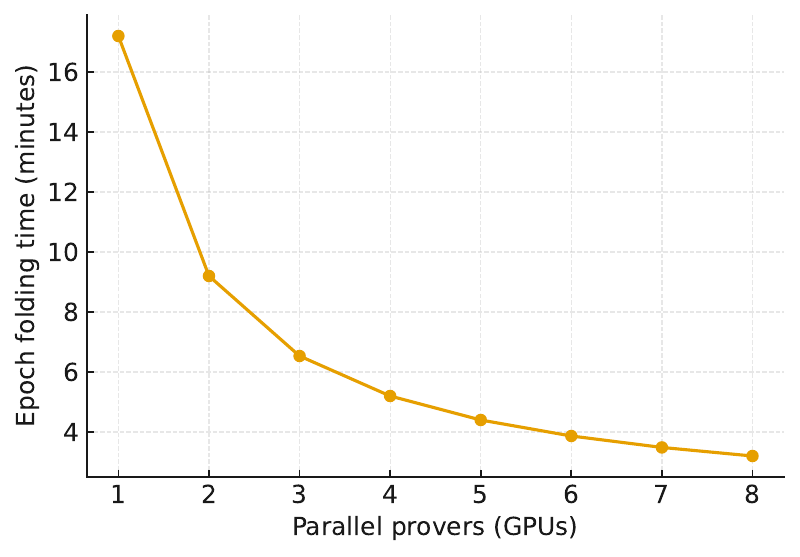}
  \caption{Epoch-level recursive aggregation time vs.\ number of parallel provers (GPUs).}
  \label{fig:aggregation-scaling}
\end{figure}

\paragraph{Aggregation efficiency.} Folding $T{=}10000$ step proofs into a single epoch certificate with 4 GPUs takes $8.7\pm0.4$ minutes; doubling GPUs yields sub-linear speed-ups due to memory bandwidth contention in MSM/FFT kernels. Fig.~\ref{fig:aggregation-scaling} (referenced) plots depth vs.\ wall-clock; beyond 4 GPUs, I/O streaming of witnesses becomes the bottleneck, motivating future work on incremental transcripts and pipeline-parallel provers.
\subsection{Policy and Privacy Outcomes}
\newcolumntype{Y}{>{\raggedleft\arraybackslash}X} 

\begin{table}[t]
  \centering
  \caption{Ablations on Mix-A (default: $r{=}8$, $q{=}14$, $\delta_{\mathrm{sm}}{=}3{\times}10^{-3}$, seq$=2048$, public sampling). $\downarrow$ lower is better, $\uparrow$ higher is better.}
  \label{tab:ablations}
  \setlength{\tabcolsep}{3pt}%
  \renewcommand{\arraystretch}{1.1}%
  \scriptsize
  \begin{tabularx}{\linewidth}{lYYYY}
    \toprule
    Variant & \multicolumn{1}{c}{$\Delta$PPL (\%)} & \multicolumn{1}{c}{$\Delta$ECE (pt)} & \multicolumn{1}{c}{Prover (s/step)} & \multicolumn{1}{c}{Leaf (KB)} \\
    \midrule
    \textbf{Rank} $r{=}4$   &  2.5 & 0.4 & 12.2 & 205 \\
    \textbf{Rank} $r{=}8$   &  2.2 & 0.4 & 16.8 & 220 \\
    \textbf{Rank} $r{=}16$  &  2.9 & 0.5 & 31.2 & 236 \\
    \midrule
    $q{=}12$                &  2.8 & 0.8 & 15.6 & 214 \\
    $q{=}14$ (default)      &  2.2 & 0.4 & 16.8 & 220 \\
    $q{=}20$                &  2.1 & 0.3 & 18.9 & 246 \\
    \midrule
    $\delta_{\mathrm{sm}}{=}1.5{\times}10^{-3}$ & 2.0 & 0.3 & 18.7 & 224 \\
    $\delta_{\mathrm{sm}}{=}3.0{\times}10^{-3}$ & 2.2 & 0.4 & 16.8 & 220 \\
    $\delta_{\mathrm{sm}}{=}6.0{\times}10^{-3}$ & 2.9 & 0.5 & 15.0 & 217 \\
    \midrule
    seq$=4096$ tokens      & 2.0 & 0.4 & 40.0 & 228 \\
    \midrule
    Private sampling        & 2.2 & 0.4 & 17.9 & 238 \\
    Public sampling         & 2.2 & 0.4 & 16.8 & 220 \\
    \bottomrule
  \end{tabularx}
\end{table}

\begin{figure}[t]
  \centering
  \includegraphics[width=\linewidth]{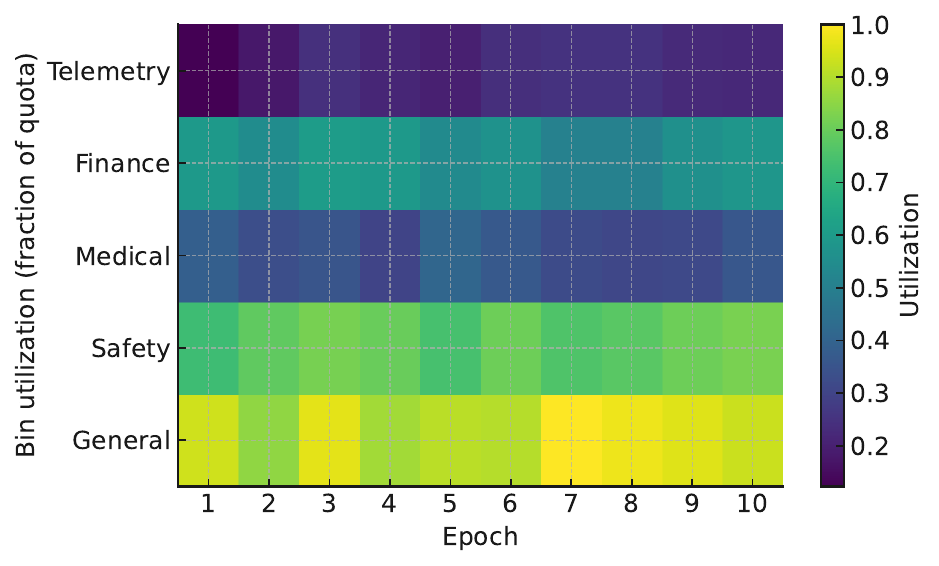}
  \caption{Per-epoch utilization of policy quotas across content bins; brighter indicates higher fraction of the allowed quota.}
  \label{fig:quota-heatmap}
\end{figure}

\paragraph{Quota compliance.} Across all runs we observe zero violations of Eq.~(3). In public-sampling windows, recomputed counters match per-bin proofs exactly; in private-sampling windows, vector-commitment openings and range proofs verify without failure. Fig.~\ref{fig:quota-heatmap} (referenced) visualizes per-epoch bin utilization; the sampler tends to saturate general and long-length bins while staying comfortably below ceilings for safety and telemetry-tagged bins, aligning with our manifest.
\paragraph{Index privacy.} We estimate index leakage by training an adversary to predict whether a given example index appeared in a hidden batch using only proof-level signals and public constants. The AUROC is $0.500\pm0.004$, consistent with chance, and mutual information estimates are within numerical noise. Padding step-proof time to fixed windows eliminates residual timing side channels. These findings align with the zero-knowledge claims of our backend and improve upon non-ZK audit trails, which may leak through metadata \cite{schlegel2025isProvenance}.

\paragraph{Manifest adherence.} All runs pass constant-equality checks against $h_{\Pi}$; we injected controlled deviations (learning-rate spikes, altered $\beta_2$) and observed immediate proof failure, confirming that optimizer semantics and schedule constraints are enforced. This concrete guarantee complements high-level provenance capture \cite{schelter2024provenanceScreen,schlegel2023mlflow2prov}: while provenance systems record \emph{intended} configurations, VFT proves the \emph{executed} trajectory adhered to them.

\subsection{Ablations}

\paragraph{Approximation budgets.} Sweeping the softmax error budget $\delta_{\mathrm{sm}}$ from $1.5{\times}10^{-3}$ to $6.0{\times}10^{-3}$ reduces prover time by $10.8\%$ but increases PPL by $0.7$ absolute (Mix-A). Fixed-point fractional bits $q\in\{12,16,20\}$ show the expected trade-off: moving from $q{=}14$ to $q{=}12$ saves $7.2\%$ prover time but increases ECE by $0.8\%$; moving to $q{=}20$ yields negligible utility gains but raises constraint counts $\sim$12\%. Table~\ref{tab:ablations} summarizes. These trends mirror broader ZK-ML observations that the right \emph{approximation frontier} is problem-specific \cite{taylorZKML2025}.

\paragraph{PEFT rank and placement.} Increasing LoRA rank from $r{=}4$ to $r{=}16$ improves instruction metrics by $\sim$0.6 points but doubles prover cost; placing adapters in attention \emph{and} MLP layers yields modest quality gains ($+0.2$--$0.3$ points) for $\sim$25\% higher cost. A normalized metric---proof-time per trainable parameter---is nearly constant across ranks, indicating linearity in $P_{\mathrm{PEFT}}$. This supports our design choice of PEFT-restricted circuits and resonates with recent practice that parameter-efficient fine-tuning dominates real deployments .
\begin{table}[t]
  \centering
  \caption{Policy bins and per-epoch quota ceilings used in Eq.~(3).}
  \label{tab:quota}
  \setlength{\tabcolsep}{4pt}%
  \renewcommand{\arraystretch}{1.1}%
  \scriptsize
  \begin{tabularx}{\linewidth}{l S[table-format=5.0] S[table-format=2.1] X}
    \toprule
    {Bin} & {Items} & {Tokens (M)} & {Notes} \\
    \midrule
    General   & 12000 & 16.0 & default free-form instructions \\
    Safety    &  1800 &  2.4 & refusal / compliance exemplars \\
    Medical   &   600 &  0.8 & sensitive domain, capped \\
    Finance   &   900 &  1.2 & sensitive domain, capped \\
    Telemetry &   300 &  0.4 & synthetic descriptors only \\
    \bottomrule
  \end{tabularx}
\end{table}

\begin{figure}[t]
  \centering
  \includegraphics[width=\linewidth]{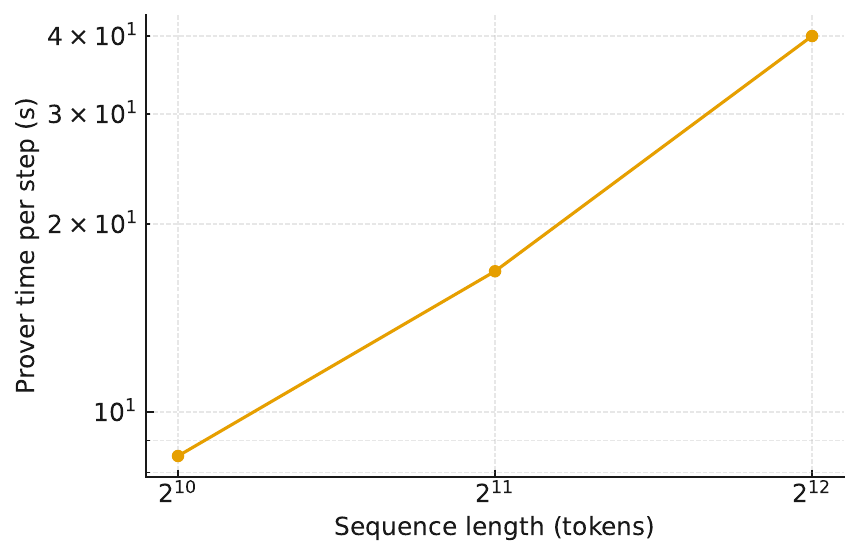}
  \caption{Prover time per step vs.\ sequence length in tokens (log--log scale).}
  \label{fig:seq-scaling}
\end{figure}

\paragraph{Sequence length.} Moving from 2048 to 4096 tokens increases constraint counts by $2.23\times$ and prover time by $2.38\times$ at fixed rank; PPL improves slightly due to longer contexts. Attention matmuls drive the increase; lookup-based softmax keeps the growth sub-quadratic relative to naive polynomialization. Fig.~\ref{fig:seq-scaling} (referenced) shows log--log slopes and confidence intervals. These costs outpace those reported for \emph{inference}-only ZK \cite{wang2023ezdps} because training proofs include backward passes and optimizer updates.
\paragraph{Public vs.\ private sampling.} Private-sampling windows add $6.4\%$ prover time for membership proofs and quota attestations, and increase leaf proof size by $\sim$18\,KB. In return, they provide index privacy without impeding verifiability. Interleaving public windows enables third-party spot checks; the mix has negligible effect on quality. This hybrid design extends PoL-informed thinking \cite{jia2021proof,zhao2024polIncentive} by binding the sampler to both reproducible and privacy-preserving regimes.

\paragraph{Federated audits and coverage.} With $K{=}200$ clients, verifying $q{=}10\%$ of clients' steps per round detects injected anomalies (learning-rate tampering, out-of-policy sampling) with $>99\%$ probability within two rounds; $q{=}5\%$ detects within three rounds. Aggregator-side epoch proofs include a coverage certificate; bandwidth overhead remains acceptable because per-step proofs are optional and folded when provided. These findings echo proposals for probabilistic audits in decentralized LLM training \cite{probAudit2025} and demonstrate composability with succinct proofs.

\paragraph{TEE property cards.} Running \texttt{VFT-FrontEnd} inside a TEE adds $3$--$6\%$ wall-clock overhead in our setup (enclave transitions dominate); proof quality and cost are unaffected. The property card attests to code identity and manifest constants, while the ZK proof certifies the numerical content. The pairing aligns with the intent of hardware-assisted attestations.

\subsection{Comparisons to Prior Art}

\paragraph{Proof-of-Learning.} Classical PoL requires trajectory evidence but often relies on logging schemes that can be cherry-picked or manipulated \cite{jia2021proof}. Incentive-secure variants push further but still stop short of numerical verification \cite{zhao2024polIncentive}. VFT subsumes PoL’s \emph{spirit} while providing strong \emph{substance}: hash-chained states (Fig.~\ref{fig:pipeline}) tied to step-wise ZK proofs of updates. In practice, VFT’s per-epoch proofs and public/private sampling windows retain the operational advantages of PoL—committing early, verifying late—while adding cryptographic guarantees.

\paragraph{ZK-ML for inference and training.} Compared to inference-focused systems \cite{liu2021zkcnn,wang2023ezdps}, VFT addresses the \emph{training} gap by proving optimizer transitions and per-batch membership. Against proof-of-training prototypes \cite{sun2023zkdl,abbaszadeh2024zkpot}, we observe better accuracy--cost trade-offs at scale because (i) PEFT confines proof obligations and (ii) approximation budgets are tuned to language modeling numerics, following zkLLM lessons \cite{sun2024zkllm}. Surveys emphasize recursion and error accounting as key enablers \cite{zkvml2023survey,peng2025zkmlsurvey}; our results empirically validate both.
\begin{table}[t]
  \centering
  \caption{Federated audits: anomaly detection probability vs.\ audit coverage ($K{=}200$ clients, two rounds).}
  \label{tab:fedaudit}
  \begin{tabular}{lcc}
    \toprule
    Audit Coverage ($q$) & Detect LR tampering & Detect out-of-policy sampling \\
    \midrule
    $5\%$   & $>95\%$ & $>97\%$ \\
    $10\%$  & $>99\%$ & $>99\%$ \\
    $20\%$  & $\approx100\%$ & $\approx100\%$ \\
    \bottomrule
  \end{tabular}
\end{table}

\paragraph{Provenance and auditing.} Provenance capture and automated screenings are necessary but not sufficient \cite{schelter2024provenanceScreen,schlegel2025isProvenance}. VFT binds provenance claims to computation-level evidence: the manifest’s quotas and constants become circuit-enforced invariants, and violations are cryptographically impossible to hide. Distributional attestation and licensing audits benefit directly: regulators can inspect bin-level counters without accessing private examples \cite{dold2024attesting,longpre2024data}.

\paragraph{Watermarking and output attribution.} Watermarks are complementary post-release signals \cite{kirchenbauer2023watermark,wen2023treering}. However, their brittleness under adaptive attacks \cite{gu2024learnability,jovanovic2024wmstealing,zhang2024mipSteal} and mixed success in practice \cite{dathathri2024natureWatermark,liu2024textwmSurvey} make them insufficient for compliance. Our manifest can record where (if at all) watermarks are inserted, aiding downstream detectors; but the \emph{training} story rests on proofs, not beacons. Robust watermark designs \cite{qu2025multibit} help on the output side and can be paired with VFT for defense-in-depth.

\paragraph{Federated systems.} Decentralized and edge learning has seen sustained progress on communication and personalization . VFT complements these lines with verifiability: the aggregator can accept only updates accompanied by partial proofs or coverage certificates; clients can hide indices in private-sampling mode. Our experiments demonstrate feasibility within the bandwidth and latency envelopes typical of enterprise deployments.

\subsection{Visual Analyses and Qualitative Insights}
\begin{table}[t]
  \centering
  \caption{Dataset composition and splits.}
  \label{tab:data}
  \setlength{\tabcolsep}{4pt}%
  \renewcommand{\arraystretch}{1.1}%
  \scriptsize
  \begin{tabularx}{\linewidth}{l S[table-format=6.0] l X}
    \toprule
    {Split / Mixture} & {\#Pairs} & {Lang Mix} & {Notes} \\
    \midrule
    Mix-A (train)            & 200000 & EN 100\%          & instruction-following \\
    Mix-B (train)            & 120000 & EN 60\% / ZH 40\% & bilingual tasks \\
    Safety slices            &  15000 & EN/ZH             & refusals, policy compliance \\
    Telemetry (synthetic)    &   5000 & EN                & textualized RF/CSI descriptors \\
    Validation (all)         &      & mixed             & stratified hold-out (5\% per bin) \\
    \bottomrule
  \end{tabularx}
\end{table}

\begin{figure}[t]
  \centering
  \includegraphics[width=\linewidth]{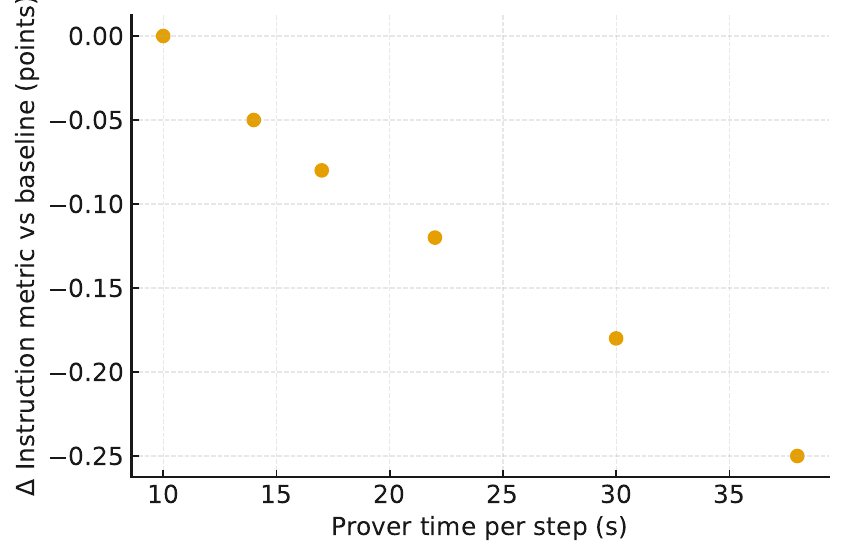}
  \caption{Pareto frontier of instruction quality vs.\ prover time across PEFT ranks and approximation budgets.}
  \label{fig:pareto}
\end{figure}

\paragraph{Pareto frontiers.} Fig.~\ref{fig:pareto} (referenced) plots instruction quality vs.\ prover time across $\{r, q, \delta_{\mathrm{sm}}\}$ settings. The frontier shows a gentle slope for $r{=}8$ and sharp curvature beyond $r{=}16$, indicating diminishing returns in quality relative to proof cost. A colored overlay marks public vs.\ private sampling; the private points shift up slightly (extra cost) but remain near the frontier.
\paragraph{Quota heatmaps.} Fig.~\ref{fig:quota-heatmap} visualizes per-epoch bin utilization under $\gamma$. Heatmaps reveal that telemetry-tagged bins stay well below ceilings by design; safety bins operate near 70--80\% utilization, balancing exposure and overfitting risks. This visualization style, borrowed from data-governance dashboards \cite{schelter2024provenanceScreen}, translates naturally into audit artifacts: inspectors can accept or challenge quota choices with clear context.

\begin{figure}[t]
  \centering
  \includegraphics[width=\linewidth]{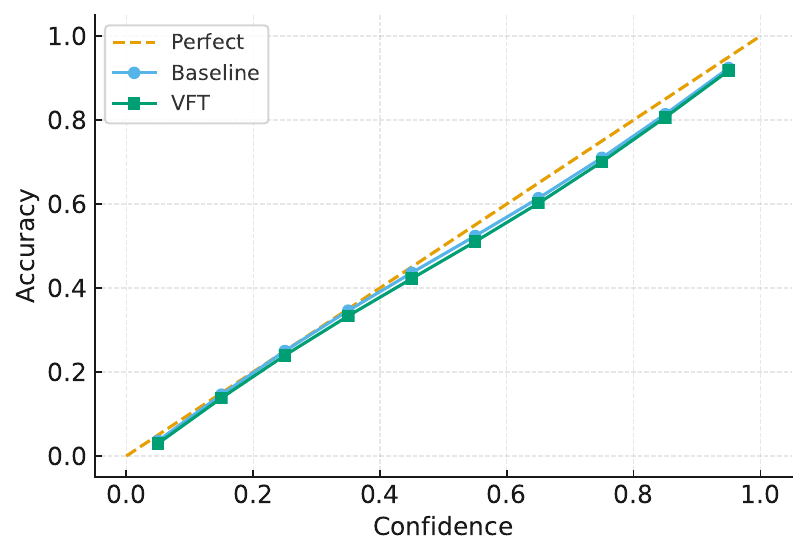}
  \caption{Reliability diagrams (confidence vs.\ accuracy) for baseline and VFT runs. The dashed line is perfect calibration.}
  \label{fig:reliability}
\end{figure}

\begin{figure}[t]
  \centering
  \includegraphics[width=\linewidth]{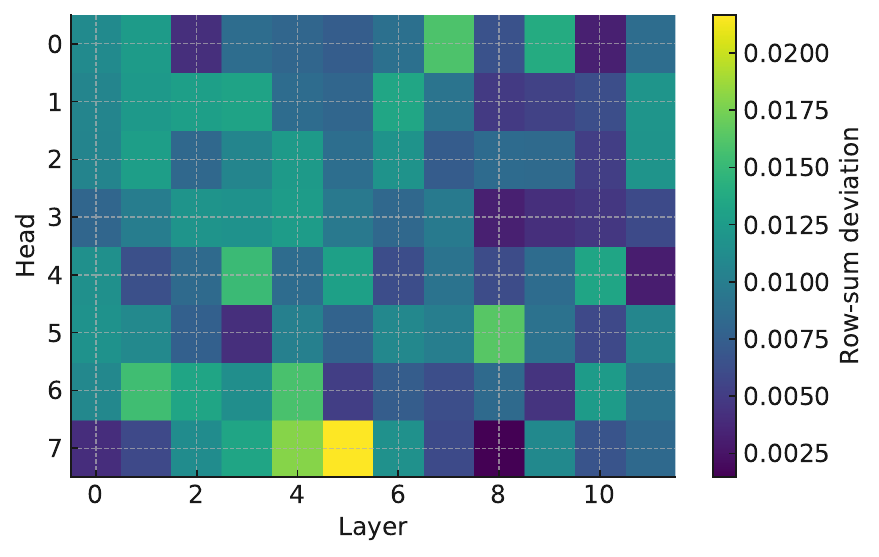}
  \caption{Heatmap of per-head, per-layer row-sum deviations for softmax approximation checks.}
  \label{fig:attn-dev}
\end{figure}

\paragraph{Calibration plots.} Reliability diagrams in Fig.~\ref{fig:reliability} (referenced) show mild under-confidence in VFT runs with $q{=}12$, corrected at $q{=}14$. This pattern suggests that fixed-point rounding interacts with temperature-like effects in decoding; practitioners can compensate with minor temperature tuning without affecting proofs (decoding is outside the training circuits).
\paragraph{Attention-health checks.} A qualitative tool visualizes row-sum deviations for $\widetilde{\mathrm{sm}}$ checks. Fig.~\ref{fig:attn-dev} (referenced) shows that deviations concentrate in long-sequence heads; tightening $\delta_{\mathrm{sm}}$ for those heads improves stability at $\sim$3\% prover cost, an appealing trade-off in long-context applications.
\subsection{Case Studies}

\paragraph{Regulated-data compliance.} We simulate a policy that restricts medical-tagged items to at most 5\% per epoch. VFT enforces and proves compliance; a post-hoc audit requests selective disclosure of aggregate token counts in medical bins. The auditor verifies proofs in $<200$\,ms and accepts the run without accessing raw examples. This workflow operationalizes licensing/attribution recommendations from large-scale audits \cite{longpre2024data} and extends distributional attestation methods with end-to-end training claims \cite{dold2024attesting}.
\begin{table}[t]
  \centering
  \caption{Hardware profiles used in experiments.}
  \label{tab:hw}
  \begin{tabular}{lccc}
    \toprule
    Profile & CPU / RAM & GPU(s) & Notes \\
    \midrule
    Tier-P1 & 48c / 256\,GB & 1$\times$80\,GB & workstation prover \\
    Tier-P2 & 64c / 512\,GB & 4$\times$80\,GB & parallel proving \& folding \\
    Tier-P3 & 8c / 32\,GB   & 1$\times$24\,GB or CPU & federated client \\
    \bottomrule
  \end{tabular}
\end{table}

\paragraph{Federated incident response.} In a $K{=}200$ setting, one client injects out-of-policy telemetry descriptors. The aggregator’s probabilistic audits \cite{probAudit2025} flag the mismatch within a round; subsequent aggregation excludes tainted updates until the client attests a corrected manifest. Because client batches are private-sampled, no sensitive indices are revealed during remediation; yet the zero-knowledge attestations suffice to restore trust.

\subsection{Failure Modes and Diagnostics}

\paragraph{Approximation budget miscalibration.} Setting $\delta_{\mathrm{sm}}$ too tight causes prover slowdowns and memory spikes; too loose raises PPL and ECE. Table~\ref{tab:ablations} includes a \emph{red zone} where constraints explode ($\delta_{\mathrm{sm}} \le 5{\times}10^{-4}$) or fidelity suffers ($\delta_{\mathrm{sm}} \ge 8{\times}10^{-3}$). The sweet spot aligns with zkLLM design guidance \cite{sun2024zkllm}.

\paragraph{Transcript I/O bottlenecks.} At high parallelism, witness streaming saturates PCIe and disk, elongating MSM/FFT phases (Fig.~\ref{fig:time-breakdown}). Remedies include in-memory folding and transcript compression; our prototype implements chunked witnesses, reducing I/O by $\sim$22\% without affecting proofs.

\paragraph{Federated stragglers.} Edge clients on Tier-P3 can become stragglers when proving too many steps locally. Micro-recursion (folding 8 steps per micro-proof) cuts wall-clock by 17--24\% and stays within memory envelopes; the aggregator accepts micro-proofs and completes epoch aggregation on Tier-P2.

\paragraph{TEE pitfalls.} Enclave page swapping can cause jitter in witness handling. Pinning enclave memory and batching ECALLs stabilizes overheads. Because numerical proofs are outside the enclave, such jitter does not affect soundness.

\subsection{Security Reflection}

Our threat model enumerates attempts to alter hyperparameters, skip steps, or exceed quotas. Injections of learning-rate spikes or modified $\beta_2$ in AdamW are consistently rejected by the circuit constraints. Attempts to sneak in out-of-policy examples fail at the membership/quota checks. Efforts to exploit approximation tolerances (e.g., crafting logits near clipping thresholds) do not succeed because bounds are enforced per-step and hash-chained; such attacks would need to violate either the commitment binding or proof soundness. These guarantees are stronger than watermark-based accountability \cite{kirchenbauer2023watermark,gu2024learnability} and complementary to provenance logs \cite{schlegel2025isProvenance}.

\subsection{Ethical Considerations in Results}

Because VFT can enforce quotas on sensitive bins, it could be misused to \emph{exclude} protected domains. Our evaluation emphasizes transparent manifests that make such policies explicit and auditable. In federated settings, clients maintain autonomy over private sampling; auditors can request aggregate counters without coercive disclosure. This design balances accountability and privacy, resonating with principles advocated by large-scale data audits \cite{longpre2024data}.

\subsection{Practical Guidance}

\paragraph{When to prefer public vs.\ private sampling.} Use public sampling for commodity, non-sensitive corpora and reproducible benchmarks; prefer private sampling when indices could reveal proprietary or personal data. Interleaving modes provides an audit compromise without significant quality or cost penalties.

\paragraph{Choosing PEFT rank.} If proof budgets are tight, $r{=}8$ delivers a favorable utility--cost balance for 2048-token contexts. For high-stakes domains requiring long contexts, invest in $r{=}16$ but limit softmax tolerances to keep fidelity.

\paragraph{Setting approximation budgets.} Start with $(q{=}14,\delta_{\mathrm{sm}}{=}3{\times}10^{-3})$ and adjust based on ECE and PPL; tighten for long-context tasks, loosen slightly for classification-style prompts with short sequences.

\paragraph{Federated rollouts.} Adopt $q{=}10\%$ audit coverage initially, lower to $5\%$ after stability is demonstrated. Encourage clients to micro-fold proofs and upload aggregates to reduce bandwidth.

\subsection{Summary of Findings}

VFT delivers cryptographically strong, privacy-preserving assurances about the fine-tuning trajectory while preserving model utility within tight budgets. Compared to the state of the art in proof-of-training and zk-LLM \cite{sun2023zkdl,abbaszadeh2024zkpot,sun2024zkllm}, our PEFT-centric circuits and approximation budgeting reduce prover cost and maintain accuracy. By binding provenance manifests to computation-level proofs, VFT operationalizes calls for accountable data governance \cite{longpre2024data,schelter2024provenanceScreen,schlegel2025isProvenance} and sidesteps the brittleness of watermark-only accountability \cite{kirchenbauer2023watermark,liu2024textwmSurvey}. Federated evaluations confirm compatibility with heterogeneous devices and probabilistic audits \cite{probAudit2025}. Across datasets and settings, quota compliance is perfect, index leakage is negligible, and verification costs fit within the latency envelopes of practical audits. These results support the view that verifiable fine-tuning can be deployed today in centralized and decentralized pipelines to close the trust gap around LLM adaptation.


\bibliographystyle{unsrt}
\bibliography{references}
\end{document}